NICK MILTON

# KNOWLEDGE

# TECHNOLOGIES

PUBLISHING STUDIES

directed by Giandomenico Sica

VOLUME 3

**N. R. MILTON**

# KNOWLEDGE TECHNOLOGIES

**Copyright and license**

You are free:
**to Share** — to copy, distribute and transmit the work
**to Remix** — to create and reproduce adaptations of the work

Under the following conditions:
**Attribution.** You must attribute the work in the manner specified by the author or licensor (but not in any way that suggests that they endorse you or your use of the work).
**Noncommercial.** You may not use this work for commercial purposes.
\* For any reuse or distribution, you must make clear to others the license terms of this work. The best way to do this is with a link to this web page.
\* Any of the above conditions can be waived if you get permission from the copyright holder.
\* Nothing in this license impairs or restricts the author's moral rights.

The work is licensed by the author through the following license:
Creative Commons license
Attribution-NonCommercial 3.0 Unported

**Note for the Reader**

In our view, doing research means building new knowledge, setting new questions, trying to find new answers, assembling and dismantling frames of interpretation of reality.

**Do you want to participate actively in our research activities?**

**Submit new questions!**

Send an email to the address **questions@polimetrica.org** and include in the message your list of questions related to the subject of this book.

Your questions can be published in the next edition of the book, together with the author's answers.

**Please do it.**

**This operation only takes you a few minutes but it is very important for us, in order to develop the contents of this research.**

Thank you very much for your help and cooperation!

We're open to discuss further collaborations and proposals.
If you have any idea, please contact us at the following address:

*Editorial office*
*POLIMETRICA*
*Corso Milano 26*
*20052 Monza MI Italy*
*Phone: ++39.039.2301829*
*E-mail: info@polimetrica.org*

**We are looking forward to getting in touch with you.**

# LIST OF QUESTIONS

# Preface

Computer systems are coming into a third age. The first age concerned data manipulation, with systems that performed mathematical operations and serial procedures. The second age saw the mass use of personal computing leading to an information revolution, hence the name Information Technology. The third age sees a further shift with new computer systems being enriched with meaning and understanding, and capable of ever more powerful reasoning processes. An era of Knowledge Technologies has begun.

A major force behind Knowledge Technologies is Artificial Intelligence (AI). Since the birth of AI just over 50 years ago, there have been a number of ups and downs, false starts, successes and failures. If nothing else, AI has proved an amazing melting pot of innovation that has brought together ideas from psychologists, logicians, linguists, computers scientists, engineers, and many other disciplines. It has spawned a number of principles, methods and tools that have reached a level of maturity and robustness required by real-world applications. Significant impact is being seen in all manner of organisations. Knowledge Technologies constitute a major part of the third age of computing.

It is high time there was a book on Knowledge Technologies that was written for people who do not have a doctorate in Artificial Intelligence or Computer Science. My aim in writing this book is to make it accessible for people with an interest in the subject, and to give them a sufficient overview to allow further exploration of the subject areas.

In doing this, I am indebted to a number of experts who have helped and guided me. I give many thanks to Steve Swallow and Clive Emberey for reading and reviewing the whole book. I give many thanks also to Gianfranco LaRocca who co-authored Chapter 3 with me. The following people also deserve special thanks for their input, inspiration and support: Nigel Shadbolt, Paul Smart, Natasha Milton and Margarita Milton.

<div align="right">
Nick Milton<br>
Nottingham, UK<br>
November 2007
</div>

# 1. INTRODUCTION

## *1.1 What are Knowledge Technologies?*

Knowledge Technologies are new computer-based techniques and tools that provide a richer and more intelligent use of Information Technology. Much of their power comes from the way they combine ideas and applications from a number of fields: Psychology, Philosophy, Artificial Intelligence, Engineering, Business Studies, Computer Science and Web Technologies.

Knowledge Technologies, as the name suggests, are about doing things with knowledge. For example:

- Identifying what knowledge is important to an organisation;

- Deciding what knowledge needs to be captured to provide an appropriate solution to a real-world problem;

- Capturing and integrating knowledge from expert practitioners and existing repositories;

- Representing and storing knowledge in ways that provide ease of access, navigation, understanding, maintenance and re-use;

- Embedding knowledge in computer systems to provide significant and definable benefits to an organisation.

Knowledge Technologies are associated with a number of subject areas that have emerged in the past 10-20 years:

- **Knowledge Engineering**. This emerged from work in Artificial Intelligence (AI). It concerns the building of computer systems that solve problems in the way humans do.

- **Knowledge Based Engineering**. This emerged from the world of Computer Aided Design (CAD). It concerns the building of computer systems that help engineers (usually design engineers) to do their jobs more efficiently.

- **Knowledge Management**. This emerged from a number of business initiatives. It concerns the use of techniques and tools to make better use of the intellectual assets in an organisation.

- **Ontological Engineering**. This emerged from Knowledge Engineering, Philosophy and Computer Science. It concerns the building of knowledge structures that allow different computer systems and repositories to understand and use each other's contents.

## *1.2 What Knowledge Technologies are there?*

**KEYWORDS:**
KNOWLEDGE TECHNOLOGIES

The Knowledge Technologies presented in this book are:

- Knowledge Based Systems (see Chapter 2)

- KBE Systems  (see Chapter 3)

- Knowledge Webs (see Chapter 4)

- Ontologies (see Chapter 5)

- Semantic Technologies (see Chapter 6)

Others technologies (not described in this book) include:

- Data Mining: The extraction of previously unknown knowledge from databases;

- Case Based Reasoning: The identification of patterns and rules in previous situations (cases) so they can be applied to solve problems in new situations;

- Intelligent Agents: Computational systems that can sense and act autonomously in a complex dynamic environment, and so realise a set of goals or tasks for which they are designed;

- Natural Language Processing: Software tools that are able to extract some meaning from paragraphs of text and do something with this extracted meaning;

- Document Management Systems: Tools that support the life cycle of electronic documents and offer facilities to manage their contents, accessibility and retrieval;

- Workflow System Management: Software that aligns and integrates an organisation's resources and capabilities with business strategies to accelerate process flow and improve productivity.

## *1.3 What is Knowledge?*

**KEYWORDS:**
KNOWLEDGE, EXPERTISE

Since all Knowledge Technologies involve the use of knowledge, we should be clear what is meant by knowledge in this context.

What is knowledge? The usual answers are variations on a number of themes:

- Knowledge is a rich and highly-structured form of information;

- Knowledge is what is needed to think like an expert;

- Knowledge is what separates experts from non-experts;

- Knowledge is what is required to perform complex tasks.

All of these are useful ways of thinking about knowledge. Another way is to think in terms of this: Knowledge is a like a machine or an engine in a person's head. What does this mean? It means that knowledge is a dynamic thing linked strongly to the context and activities in which it is used. It acts like a machine that takes in data and information at one end and spurts out decisions and actions at the other end. This idea can be put into a definition:

$$\text{Knowledge is the} \left\{ \begin{array}{c} \text{ability} \\ \text{skill} \\ \text{expertise} \end{array} \right\} \text{to} \left\{ \begin{array}{c} \text{manipulate} \\ \text{transform} \\ \text{create} \end{array} \right\} \left\{ \begin{array}{c} \text{data} \\ \text{information} \\ \text{ideas} \end{array} \right\} \text{to} \left\{ \begin{array}{c} \text{perform skilfully} \\ \text{make decisions} \\ \text{solve problems} \end{array} \right\}$$

Just like a real machine, the knowledge machine in a person's head can only be completely understood if you know two things: (i) How it is **structured**, i.e. what components it is made from, and the ways they are linked together; (ii) How it **operates**, i.e. the ways in which the components behave and the processes that are happening. This is an important way in which we can view knowledge, as being about its structural components or about the processes that operate. You will see this idea appearing in a number of places in this book.

## *1.4 How is knowledge stored and represented?*

**KEYWORDS:**
KNOWLEDGE BASE, CONCEPTS, ATTRIBUTES, VALUES, RELATIONS, KNOWLEDGE REPRESENTATION, SEMANTIC NETWORK, FRAMES, LOGIC, PREDICATES

### 1.4.1 Knowledge Base

Most Knowledge Technologies (certainly all of the ones in this book) rely to a greater or lesser extent on a way of storing knowledge called a **knowledge base**. So what is a knowledge base? It is special database that holds information representing the expertise of a particular domain. To do this it is designed to have a structure that is similar to the structures that underlie human expertise. Psychologists have found that this is based on 4 main components:

- **Concepts**: The things in a domain, such as physical entities, people, documents, organisations, tasks, ideas, places, etc.

- **Attributes**: The general properties of concepts, such as colour, weight, age, length, number of employees, difficulty, etc.

- **Values**: The specific properties of something that distinguishes it from other things, such as green, heavy, 47 years, long, easy, etc.

- **Relations**: The ways in which concepts are associated with one another. Some examples are:

    ◦ *is a*, e.g. 'cat – *is a* – animal', 'PCPACK – *is a* – software tool'
    ◦ *part of*, e.g. 'leg – *part of* – cat', 'chapter 1 – *part of* – book'

- ◦ *requires*, e.g. 'animal – *requires* – food', 'organisation – *requires* – customers'
- ◦ *performs*, e.g. 'knowledge engineer – *performs* – interview expert'
- ◦ *causes*, e.g. 'rising water levels – *causes* – flood'

1.4.2 Knowledge Base

The structure of a knowledge base is determined by the purpose of the knowledge and the way it should be represented both to other IT systems and to humans. There are two basic representational formats: one based on **relations**, the other on **attributes** and values. The first can be visualised as a network of concepts with links between them, each link representing a relation. This is the basic format of a diagram called a concept map (see Section 4.2.2.2) and is also the basis of most logic languages (see Section 5.5.1). Figure 10 (near the end of the book) shows this kind of representation.

The other format, based on attributes and values, is best visualised as a frame, as shown in Figure 1.

| Coffee | |
|---|---|
| colour | brown |
| cost | medium cost |
| serving temp | hot |
| transparency | opaque |
| amount of alcohol | no alcohol |
| amount of milk | no/some milk |

| Vodka | |
|---|---|
| colour | colourless |
| cost | high cost |
| serving temp | cold |
| transparency | transparent |
| amount of alcohol | highly alcoholic |
| amount of milk | no milk |

Figure 1. Comparing the properties of two drinks using two frames

Some knowledge bases use a purely relational format, others use frames, and others combine the two. You will see this idea re-appearing at different places in the book.

## *1.5 How can Knowledge Technologies benefit an organisation?*

**KEYWORDS:**
KNOWLEDGE SHARING, TASK AUTOMATION, SYSTEM INTEGRATION, INFORMATION STORAGE, PROCESS IMPROVEMENT

Knowledge Technologies can provide a wide range of benefits to an organisation. The subsequent chapters in this book each include a section on the uses of the Knowledge Technology in question. To give a flavour of what can be achieved, Table 1 lists a number of uses and benefits.

Table 1. Uses and Benefits of Knowledge Technologies

| Uses | Knowledge Technologies | Benefits |
|------|------------------------|----------|
| Capturing and sharing expertise around an organisation | Knowledge Webs KBE Systems | Improving efficiency by providing non-experts with the expertise to do a better job |
| Automating tasks normally performed by human experts | Knowledge Based Systems KBE Systems Semantic Technologies | Reducing the cost and time involved in business processes, and allowing experts to be more effective |
| Integrating different databases and computer systems | Ontologies Semantic Technologies | Reducing the inefficiencies created by having multiple databases and IT systems that do not talk to each other |
| Capturing and storing knowledge for future uses | Knowledge Webs | Stopping the loss of knowledge when experts leave the organisation, and making better use of lessons learned |
| Providing intelligent access to information stored in databases and web sites | Semantic Technologies Ontologies | Making effective use of IT and web resources by filtering and presenting information in user-specific ways |

# 2. KNOWLEDGE BASED SYSTEMS

## *2.1 What is a Knowledge Based System?*

A Knowledge Based System (or KBS for short) is a computer program that uses Artificial Intelligence (AI) techniques to solve complex problems that would normally be performed by a person with specific expertise. Because of this, Knowledge Based Systems are also referred to as Expert Systems.

To illustrate the distinguishing features of a Knowledge Based System, as compared to other computer applications, let us look at an example.

### 2.1.1 Example – MYCIN

In the 1970s, AI researchers developed one of the first and most well-known Knowledge Based Systems, called MYCIN [1]. MYCIN was able to diagnose and recommend treatments for blood infections. To do this it required plenty of knowledge that was acquired from expert physicians. This expertise was represented in a knowledge base, mostly in the form of IF-THEN rules. As an example, here is an English version of a MYCIN rule:

> IF the stain of the organism is gram positive AND the morphology of the organism is coccus AND the growth formation is clumps THEN there is suggestive evidence (0.7) that the identity of the organism is staphylococcus.

Note, that this rule combines a number of factors (with logical AND operators) so that if all are true then the conclusion is true with a probability factor of 0.7. Use of this kind of statistical reasoning allows a richer and more human-like way of solving problems than other approaches. For example, it can provide intelligent guesses when a limited number of facts are available.

## 2.1.2 Developments to MYCIN

Later developments to MYCIN illustrate some other key features of Knowledge Based Systems.

- A program called TEIRESIAS was written to provide a more intelligent user interface to MYCIN. This allowed the system to explain how it arrived at its conclusions to the user (a doctor). This was achieved by inspecting the particular rules that had been applied to the case in question. For example, if the doctor wanted to know why an organism had been classified as staphylococcus, then the rule shown above could be used to say why, i.e. that the stain was gram positive, the morphology was coccus and the growth formation was clumps.

- A new Knowledge Based System called NEOMYCIN was created that had a richer knowledge base and used meta-rules that represented strategic knowledge. So rather than use a simple rule-base, NEOMYCIN used a taxonomy (tree) of diseases going from general to specific and would tackle a problem by navigating through the tree in a strategic fashion. In this way, it could ask more appropriate questions to the user to decide between candidate diseases.

• A so-called "expert system shell" called EMYCIN was developed that used all of the structure and general code of MYCIN but none of its domain-specific content. This allowed other knowledge engineers to quickly develop new Knowledge Based Systems, since they had only to populate it with the specific domain content. (As it happens, this turned out to be easier said than done – more of this later).

### 2.1.3 General Features of a Knowledge Based System

MYCIN and its descendents illustrate some of the major features of a general Knowledge Based System:

• It incorporates a substantial knowledge base containing information, structures and rules normally held in an expert's head;

• It is programmed to solve problems in a similar way to that of an expert practitioner, e.g. make inferences based on the case at hand and adopt the right strategy to solve the problem;

• It can deal with incomplete information by making requests for further information or by making intelligent guesses at the answer (just as an expert has to do when there is limited information);

• It has a user interface that makes intelligent requests for information and can explain how it has arrived at its answers;

• Its development can be made cost-effective by re-using generic structures, rules and problem-solving methods.

## *2.2 How does a Knowledge Based System operate?*

**KEYWORDS:**
KNOWLEDGE BASE, WORKING MEMORY, INFERENCE ENGINE, RULES, PROBLEM SOLVING MODELS

The architecture of a basic Knowledge Based System is shown in Figure 2:

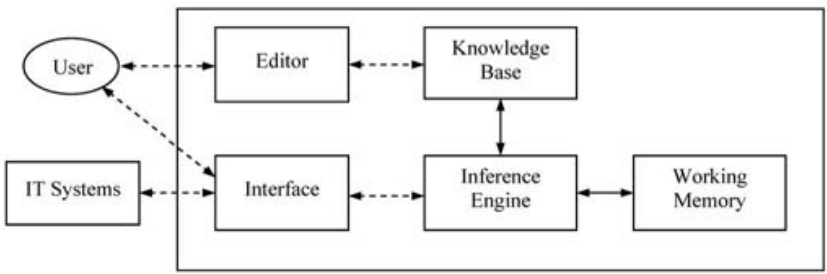

Figure 2. Basic architecture of a Knowledge Based System

The main components of a Knowledge Based System are:

1. A **knowledge base** that contains domain-specific information, structures and rules;

2. A **working memory** (also known as a "blackboard") which holds case-specific data (e.g. facts about the initial problem and intermediate results);

3. An **inference engine** (or reasoning engine) that controls and directs the solving of problems by making inferences, i.e. it uses the knowledge base to alter the contents of the working memory.

In addition, most modern systems have 2 further elements:

1. An **interface** to other computer systems and/or to human users;

2. An **editor** that allows a knowledge engineer or domain expert to inspect and update the knowledge base.

To illustrate how a Knowledge Based System operates, let us look at a very simple example.

2.2.1 Example

In this example, the system has a knowledge base containing 3 rules:

>   Rule 1: IF flow pressure>20 AND flow is rippling
>   THEN flow is vortex
>
>   Rule 2: IF core temp<90 AND flow is unstable
>   THEN open the U-valve
>
>   Rule 3: IF P-valve is open OR flow is vortex
>   THEN flow is unstable

Let us suppose, the initial contents of working memory are:

- flow pressure = 30
- core temp = 80
- flow is rippling
- P-valve is closed

The reasoning would operate as follow:

- On the first cycle, Rule 1 matches the contents of working memory, so 'flow is vortex' is added to working memory;

- On the next cycle, Rule 3 matches (because 'flow is vortex' is in working memory), so 'flow is unstable' is

added to working memory;

- On the next cycle, Rule 2 matches, so 'open the U-valve' is added to working memory. Presumably this would trigger an automatic action or a message for the user to take an action.

### 2.2.2 Control and Strategy

A major aspect of a Knowledge Based System is how to control the reasoning. For example, what happens if more than one rule matches? Should all the rules be used, and if so, in what order?

There are a number of aspects to this:

- Controlling the way in which rules are matched;
- Controlling which rules are used by the inference engine;
- Controlling the order in which rules are used by the inference engine.

Let us explore some of this in more detail.

### 2.2.3 Chaining

There are two ways in which rules can be matched:

- The inference engine can look for matches using the left-hand side of each rule (the part before THEN). In this case, the right-hand side is added to the working memory.

- Alternatively, the inference engine can look for matches using the right-hand side of each rule (the part after THEN). In this case, the left-hand side is added to working memory.

These two methods give rise to two types of reasoning: forward chaining and backward chaining.

- Forward chaining is the type of reasoning used in the example above. The contents of the working memory, which are initially the case data, are matched to the left-hand side of the rules, and this process continues until a conclusion or answer is reached.

- Backward chaining, as the name suggests, operates in the opposite direction to forward chaining, i.e. you start with a conclusion (a guess at what the answer might be) then match the right-hand side of rules to see if the left-hand sides are present in the working memory. By using this approach, you can propose some candidate solutions and eliminate those that do not fit the facts.

For most cases, forward chaining is good for solving problems when the details of the answer are unknown, e.g. when creating a plan or designing a product. Backward chaining is often better, when the problem is to select from a range of candidate answers, e.g. when classifying something as one type or another.

2.2.4 Problem Solving Models

A very important aspect of Knowledge Based Systems is to break a problem into chunks, so that the inference engine works on one part of the task (and one part of the knowledge base) at any one time. This helps to control the reasoning process. It is also of huge value if these chunks are generic, i.e. they apply to a whole set of Knowledge Based Systems rather than just one. In this way the reasoning process can be controlled using a standard and re-usable structure. One way to achieve this is by having a library of general Problem

Solving Models, giving the following advantages:

- They make the development of a Knowledge Based System much easier and more efficient because the reasoning process can be selected from a library of standard models rather than writing it from scratch;

- The way in which the knowledge base is structured can also be standardised making it easier to use, update and maintain;

- The Knowledge Based System becomes re-usable for other projects of the same type by just altering the knowledge base and little else.

A major feature of the CommonKADS methodology is a library of Problem Solving Models (see the next section).

*2.3 How is a Knowledge Based System developed?*

**KEYWORDS:**
SYSTEM DEVELOPMENT, COMMONKADS, PROBLEM SOLVING MODELS

Most modern Knowledge Based Systems are developed in 2 main stages:

1. A knowledge acquisition stage in which a knowledge engineer captures and models knowledge from a domain expert to build a knowledge base [2]. This stage also includes phases of: (i) defining the system, and (ii) specifying what knowledge will be captured from the experts (called scoping).

2. An implementation stage in which a programmer takes the information and knowledge base and programs this into the system. This stage also includes phases of test and refinement.

The most comprehensive and sophisticated methodology is CommonKADS [3], which adopts a spiral approach, rather than a simple 2-stage approach. It includes activities involving organisational analysis, knowledge acquisition, knowledge analysis, knowledge modelling, capture of user requirements, analysis of system integration issues, design of the Knowledge Based System, and project management.

A CommonKADS project uses a model-driven approach, i.e. a number of models are constructed. This happens across 3 phases of work as described below.

### 2.3.1 Phase 1: Organisation Model, Task Model and Agent Model

During this phase, a number of questions are answered: Why is a Knowledge Based System required? What problems will be addressed by the solution? What are the benefits and costs? What are the impacts on the organisation?

To address these questions, the context and environment into which the system will be developed and implemented must be fully understood. To do this, three things must be analysed: (i) the organisation, (ii) the relevant tasks, (iii) the agents (people and computers) that perform the tasks. This is done by creating models in each of these areas, in the form of a set of charts and tables:

- The Organisation Model includes the knowledge-oriented problems and opportunities, a breakdown of the business process and the knowledge assets involved to create a feasibility decision document;

- The Task Model maps out the relevant tasks and their features;

- The Agent Model includes the agents, the tasks they perform, the knowledge employed for the tasks (in-

cluding knowledge bottlenecks) and an assessment of how the Knowledge Based System would impact and change people's roles.

### 2.3.2 Phase 2: Knowledge Model and Communication Model

The second phase of the project addresses these questions: What is the nature and structure of the knowledge involved? What is the nature and structure of the corresponding communication?

To answer these questions, a conceptual description of the knowledge applied in a task is created. This is achieved by building a Knowledge Model and a Communication Model.

The Knowledge Model is an implementation-independent description of the knowledge components involved in carrying out a task.

To help create the Knowledge Model, a task template is selected and then populated with domain knowledge. The task template is based on a Problem Solving Model (see Section 2.2.4) selected from the CommonKADS library of Problem Solving Models. For example, one of these is the **Diagnosis model**, which describes how an expert finds a fault in a system, whether this is a doctor diagnosing a disease, a car mechanic finding the cause of an oil leak or a programmer investigating a software bug.

As well as Diagnosis, the other main Problem Solving Models in CommonKADS are:

- **Classification** - Establishing the correct category for an object;

- **Assessment** - Finding a decision category for a case based on a set of domain-specific norms;

- **Monitoring** - Analysing an ongoing process to find out whether it behaves according to expectations;

- **Synthesis** - Designing a structure that fulfils a set of requirements;

- **Configuration Design** - Finding an assembly of components that satisfies a set of requirements and obeys all the constraints;

- **Assignment** - Creating a relation between two groups of objects that satisfies the requirements and obeys all the constraints;

- **Planning** - Generating a plan consisting of a ordered set of activities that meet a goal or set of goals;

- **Scheduling** - Creating a schedule of temporally-sequenced activities.

Each Problem Solving Model is used to drive the elicitation, analysis and modelling activities required to build the Knowledge Model.

Alongside the Knowledge Model, a Communication Model is constructed, which is a description of the interactions between the various agents involved in a task. This is achieved in a conceptual and implementation-independent way.

### 2.3.3 Phase 3: System Design

The third phase of the project addresses these questions: How should the knowledge be implemented in a computer system? What software architecture and computational mechanisms should be used?

The models created in the previous phases constitute a requirements specification for the Knowledge Based System, broken down into different aspects. Based on these, the Design Model is created. This gives the technical system specification in terms of architecture, platform, software modules, representational constructs and computational mechanisms.

The Design Model is used to program the Knowledge Based System. If a previous system uses the same Problem Solving Model, then much of the code can be re-used in the new system, thus saving time and effort.

## *2.4 What are the uses of Knowledge Based Systems?*

**KEYWORDS:**

TASK AUTOMATION, PROCESS IMPROVEMENT

Knowledge Based Systems are an ideal solution:

- If you need a computer system to replace or support some of the problem-solving tasks of a human expert. This will be the case if it is expensive, inefficient, risky or dangerous for a human to do the task.

- If it is expensive, inefficient, difficult or impossible for another type of computer system to do the task. Although expertise can be implemented in conventional programming approaches, considerable effort is required. This is because human experts solve complex problems using abstract, symbolic approaches, which are easier to emulate using programming techniques based on the principles and techniques of Knowledge Engineering.

Many different types of Knowledge Based Systems can be built. The different Problem Solving Models described in the previous section mirror the different systems:

- Classification systems, e.g. a system that classifies the minerals in a rock;

- Assessment systems, e.g. a system that decides whether a person gets a loan or not;

- Monitoring systems, e.g. a system that monitors an industrial plant;

- Synthesis systems, e.g. a system that designs part of an aircraft wing;

- Configuration systems, e.g. a system that configures a computer system;

- Assignment systems, e.g. a system that assigns airplanes to airport gates;

- Planning systems, e.g. a system that plans the therapeutic actions for treating a disease;

- Scheduling systems, e.g. a system that produces a production schedule for a factory.

Thus, Knowledge Based Systems can be used in all manner of applications to do with interpreting, diagnosing, designing, debugging, repairing, instructing, controlling, planning, etc.

Knowledge Based Systems are also useful:

- When another type of computer system would require a very bespoke solution that cannot easily be re-worked and re-used;

- When the knowledge might be tacit knowledge (deep inside an expert's head) so it is unclear initially how to solve the problem, i.e. you need to take a knowledge engineering approach to discover how to solve the problem;

- When you need to be able to structure and maintain the knowledge separately from the application, e.g. when the knowledge may change and go out of date;

- When an explanation needs to be given of how the answers were arrived at;

- When a human expert does not have the time to do all of the tasks (e.g. in a military or disaster situation) and the requirement is for an intelligent system to take over some of the activities, such as decision-making;

- When incomplete information may be available but you still need an answer that is an intelligent best guess.

## *2.5 What tools and technologies are available?*

**KEYWORDS:**
PCPACK, PROTÉGÉ, CLIPS

Tools and technologies can be split into 3 types: (i) tools used to capture and model knowledge from domain experts; (ii) technologies used for the knowledge base, (iii) programming languages and tools.

Capture and modelling tools include:

- **PCPACK**: a knowledge acquisition toolkit (see Section 2.5.1 below);

- **Protégé**: a knowledge modelling and structuring tool (see Section 5.5.2).

Knowledge base technologies include:

- **XML** (see Section 4.5.1);

- **Triple Stores** (see Section 6.5.3);

- A traditional database or just text files.

Programming languages and tools include:

- Languages specifically designed for rule-based systems, such as **CLIPS** (see Section 2.5.2), **PopLog**, **Ops5 and Jess**;

- **Functional Programming Languages** such as **Lisp** and **Prolog**, useful for building the sort of components required in a Knowledge Based System;

- Almost any modern programming language can be used to write the inference engine, such as Visual Basic (VB) or Object Pascal;

- Special programming tools known as 'shells'. These became available in the 1980s and provide a template-driven approach. However, they are less popular now, since they constrain the type of system that can be built.

### 2.5.1 PCPACK

Knowledge engineers use PCPACK [4] to make the process of acquiring, modelling and storing knowledge more efficient and less prone to errors. It is a comprehensive suite of tools that allows the user to create, inspect and edit an XML knowledge base. Each tool provides a different way of viewing the knowledge base. Many types of knowledge representations are used, and a range of different capture techniques is supported. This versatility makes it one of the main support tools for CommonKADS [3]. PCPACK includes the following tools:

- Protocol tool: this allows the marking-up of interview transcripts, notes and documentation (protocols) to identify and classify knowledge elements to be added to the knowledge base;

- Ladder tool: this facilitates the creation of hierarchies of knowledge elements such as concepts, attributes, processes and requirements;

- Diagram tool: this allows the user to construct compact

networks of relations between knowledge elements, such as process maps, concept maps and state-transition diagrams;

- Matrix tool: this allows grids to be created and edited that show relations and attributes of knowledge elements;

- Annotation tool: this allows sophisticated annotations to be created using dynamic html, which include automatically generated hyperlinks to other resources in the knowledge base;

- Publisher tool: this allows a web site, or other information resource, to be created from the knowledge base using a template-driven approach that maximises reuse and standardisation (see Section 4.5.4).

### 2.5.2 CLIPS

CLIPS [5] provides a complete environment for constructing rule-based and/or object-based Knowledge Based Systems. It was first released in the mid-1980s with a forward-chaining, rule-based approach. In the 1990s, two new programming paradigms were added:

- Procedural programming: This provides CLIPS with capabilities similar to languages such as C, Java, Ada, and LISP;

- Object-oriented programming: This allows complex systems to be modelled as modular components, which can easily be reused to model other systems or to create new components.

The CLIPS shell provides the basic elements of a Knowledge Based System:

- A working memory, which consists of a fact-list and instance-list;

- A knowledge-base, which contains all the rules;

- An inference engine for controlling the execution of rules.

CLIPS is an attractive environment to use because of its low cost (in fact, it is free for non-commercial use, hence it is used by many students and academics). Other benefits are its portability, its extensibility and its capabilities. On the downside, some real-time systems that have rapid changes of input data and/or require responses every second cannot be implemented in CLIPS as its algorithm cannot provide responses so quickly. Hence, one way to write a real-time Knowledge Based System is to build a prototype in CLIPS (which is succinct and represents the rules very clearly) and then convert this into a compiled system (written in a modern programming language) that runs more efficiently and quickly.

2.5.3 Technologies in Modern Knowledge Based Systems

Most modern Knowledge Based Systems:

- Adopt an object-oriented programming approach to knowledge representation;

- Are complex systems with multiple knowledge sources, multiple lines of reasoning, and fuzzy information;

- Use multiple knowledge bases;

- Are based on proven techniques and tools for acquiring knowledge;

- Exploit the larger storage capabilities and faster processing of modern computers;

• Use web technologies to disseminate software and expertise.

## *2.6 What are the issues of Knowledge Based Systems?*

**KEYWORDS:**
ISSUES, DEVELOPMENT COSTS, BUSINESS BENEFITS

Developing a Knowledge Based System is not a cheap or trivial task. A project that does not progress using the best principles and practices will run the risk of being too expensive and not providing the impact that was envisaged. There are two areas where best practices can address the risks:

1. Minimising the development costs;

2. Maximising the impact.

Let us see how these can be achieved.

### 2.6.1 Minimising Development Costs

The knowledge engineering community has devoted much effort to minimising the time, effort and cost required to build a Knowledge Based System. One of the main achievements has been the use of generic models that can be re-used from project to project (see Section 2.2.4). In this way, a project need not start with a blank sheet of paper, but has something to populate, whether it is a shell, a skeleton, a template, an ontology or a generic model. A key innovation is the use of generic rules in the system, so that only the domain classes, relationships and attributes need to be captured in the knowledge base (rather than all the rules for performing the problem solving).

Another important development has been the increased efficiency in acquiring knowledge from domain experts. This has been achieved in various ways:

- A range of special techniques has been developed for use with experts [2], including techniques such as semi-structured interviews, laddering, concept mapping, commentary, card sorting, and repertory grid. A key to this approach is to understand what knowledge is being acquired and to select the appropriate technique.

- Knowledge acquisition software makes it much easier and quicker to construct a knowledge base. Tools that provide multiple ways of representing knowledge (such as PCPACK - see Section 2.5.1) make the process significantly more effective. They are especially good at improving the validation process, i.e. checking that the knowledge is correct, complete, consistent and relevant.

- Generic models are used to show what types of knowledge need to be acquired and what reasoning processes might be operating. Prime amongst these are the Problems Solving Models of CommonKADS (see Section 2.3.2).

Three key benefits of all this are: (i) a reduction in the time required to create a knowledge base; (ii) a reduction in the disruption of the project to the normal running of the business, i.e. experts are not away from their work for long periods of time; (iii) knowledge collected on one project can be re-used on future projects.

2.6.2 Maximising the impact

It is vital that a Knowledge Based System makes major improvements to the organisation. To ensure this is the case, it is vital that:

- There is a well-defined phase at the start of a project to investigate and analyse the opportunities for a Knowledge Based System and specify how the system will fit into the new ways of working;

- There is a well-defined scoping phase to define exactly what knowledge will be collected and where it will be captured from;

- The knowledge is captured from domain experts in ways that result in a system that performs as specified;

- The system is built using a modular approach with generic rules and is transparent, i.e. the inner workings can be examined, analysed and edited;

- The system can explain how it has arrived at its answers;

- The system is maintainable, i.e. can easily be updated as things change in the organisation, e.g. new technologies, new practices, new regulations, etc.

The system should address one or more of the following benefits:

- Decision Making: Decreased decision-making time, improved decision-making process and improved decision quality;

- Processes: Increased process and product quality, flexibility, increased output and increased productivity;

- Production: Reduced downtime, easier equipment operation, operation in hazardous environments and elimination of the need for expensive equipment;

- Information: Capture of scarce expertise, accessibility to knowledge and help desks, ability to work with in-

complete or uncertain information, knowledge transfer to remote locations and integration of several experts' opinions;

- Skills and Capabilities: Ability to provide training, ability to solve complex problems and increased capabilities of other computerized systems.

## *2.7 Where can I get more information?*

See references 1-5 in the Bibliography (Section 8).

Other sources of information:

# 3. KBE SYSTEMS

## 3.1 What is a KBE System?

A KBE (Knowledge Based Engineering) System can be considered as a special kind of Knowledge Based System endowed with advanced data processing and CAD (Computer Aided Design) capabilities. Similarly to Knowledge Based Systems, KBE systems are able to store and reuse the knowledge of experts to solve problems. However, they target the field of engineering, where solving problems by reasoning must be supplemented with analyses and computations. On top of that, the field of design engineering comes with the challenges of geometry manipulation and product configuration. In this sense, KBE is a technology able to merge the capabilities of conventional Knowledge Based Systems with those of computer aided analysis and design systems (CAE and CAD systems).

KBE Systems allow people to write dedicated programs (called KBE applications) that can perform complex and specific engineering activities more efficiently and effectively than engineers can achieve.

For example, it often takes a team of design engineers many weeks to design a complex component including the generation of geometry and the associated analysis process. With the help of a dedicated KBE application, a single engineer could achieve this same task in just a few days. The resulting savings on manpower, costs and time can be substantial. All of the time normally "wasted" by designers in lengthy and repetitive tasks can instead be dedicated to crea-tivity and innovation. This enhances worker satisfaction and product quality.

The main use of KBE is currently to support and improve the design of complex (mechanical) systems. It is not by chance then that aircraft and car manufactures are the main users of KBE technology.

KBE Systems are required because of the increasingly complex nature of engineering processes from one side, and the increasing demand for shorter and cheaper product development on the other. Engineering companies face many challenges:

- There is an increasing number of requirements, from different disciplines, that have to be fulfilled;

- The knowledge required to solve complex design problems is becoming ever more specialised hence less readily available;

- There are greater demands to produce better products in shorter time and with lower resources (to survive in the global market place);

- There are less opportunities to transfer knowledge;

- Technologies are constantly changing.

Traditional CAD (Computer Aided Design) systems are a valuable tool for designers. Indeed, they represent the most widely used computer-aided tools in design engineering. However, CAD systems model the design by means of very simple geometry features (e.g. points, lines and shapes), which are rather inadequate to support the level of design abstraction that is required, especially in the early design phase [6]. The typical CAD approach of generating complex product configurations by means of mouse gimmicks and lengthy menu clicking sessions is inadequate to support the design automation intended with KBE.

KBE Systems are neither an alternative to CAD (though they can include or use a CAD engine), nor are they classical Knowledge Based Systems (though they offer the tools to store knowledge and reason upon that to solve problems). In fact, they are a sophisticated blend of AI, CAD and Object-Oriented modelling that can alleviate the problems described above by capturing and efficiently reusing the knowledge of engineering design experts.

## *3.2 How does a KBE System operate?*

**KEYWORDS:**
PRODUCT MODEL, ADDET, MOB

A typical KBE System provides a KBE developer with:

- A programming environment to code the experts' knowledge about the design of a given product or component (i.e. how the product is defined, and the process of generating a product instantiation by the systematic application of logical rules and various algorithms and procedures);

- A browsing interface to visualise the geometry of the given product and make queries about its geometric and non-geometric attributes (e.g. size, weight, cost etc.).

Every time the KBE tool is asked to generate a new product configuration or compute some product characteristics, there is a language compiler/interpreter that executes all the programmed rules and algorithms and generates the required results. In KBE parlance, this set of programmed rules and algorithms is called a **product model** and represents the core of each KBE application.

As discussed above, an essential component of a KBE system is a CAD engine, which can be fully driven by the product model. For example, the product model can contain commands to ask the CAD engine to generate a given curve or to perform an intersection between two solids. The results of these operations can be used by the product model to make decisions about the generation of extra or different product components, or to re-compute some property of the product under consideration. To do this, it would use rules, such as these:

> IF flange_diameter > 50mm THEN number_of_holes = flange_diameter/5;
>
> IF force > 50 Newton THEN material = steel AND Number_of_supports = 6.

The basic steps in the operation of a typical KBE application [6] are:

1. Assign values to the parameters used in the definition of the product model, either by reading them from an input file (edited by the designer or automatically generated by some software tool) or by inserting them via a user interface;

2. Instantiate the product model with this set of parameter values;

3. The product model applies its body of rules and algorithms (i.e. systematically applies all of the relevant knowledge as a human expert would do);

4. The results produced by the KBE application can be visualized/queried via the graphical interface. These results might consist of a geometry model or some other models possibly not including geometrical entities.

To illustrate how a real KBE System operates, two case studies are briefly described below.

3.2.1 Case Study 1 - ADDET

This case study is about the ADDET KBE application created by Brent Vermeulen of Stork Fokker [7].

ADDET was developed to aid the design of fuselage panels for aircraft. Typically, the number of requirements and constraints in designing such a structural assembly is so high that even an expert designer struggles to find an appropriate solution. In some cases, there might not be a solution that fully satisfies the set of requirements and constraints. In other cases, more than one possible solution exists and an optimal solution must be found. ADDET allows the systematic generation of many possible solutions for panel design and searches for the best. It also warns the designer of any constraints that may have been violated or of any objectives that may not have been met. According to Vermeulen, ADDET can find 15 solutions in the time required by the expert to manually generate one solution; and ADDET's solution is often better than the expert's one.

ADDET uses the CATIA V5 CAD system as the graphical modeller and Visual Basic (VB) as the programming language to define the product model. The product model has being developed by capturing the design knowledge of Stork experts (i.e. how to model a fuselage panel, how to choose shape and positioning of various structural components, what constraints are to be respected and how to judge the quality of a given design).

These are the basic steps during ADDET operations:

1. The user (a fuselage panel designer in this case) provides input data to the product model (e.g. he/she provides the definition of the fuselage shape and panel contour);

2. A first panel design solution is automatically generated (using the product model rules written in VB and the

CATIA geometric primitives);

3. The quality of the panel model is automatically analysed using mathematical functions (objective functions and penalty functions) coded in the product model;

4. The compliance with requirements is automatically checked using an evaluation algorithm coded in the product model;

5. If compliance is not met, the solution domain is further searched and an improved panel design is automatically generated;

6. The best solution is finally exported to the user for further evaluation and possibly manual modification.

### 3.2.2 Case Study 2 - MOB

This case study is about the KBE system developed by La Rocca within the European project MOB. The project concerned the development of a computational system to support distributed, multidisciplinary design and optimisation of a blended wing body aircraft [8]. A large amount of analysis work was required to understand the quality of this novel configuration, which involved the extensive use of many different analysis tools, ranging from aerodynamics to structures. In addition, the blended wing body configuration is intrinsically very complex; any modification on one part of the aircraft has an impact on the rest of the design, which needs to be quantified. A concurrent, multidisciplinary design approach was required, which involved experts and analysis tools located in different countries and companies. The ICAD system was used to develop a dedicated KBE application, called Multi Model Generator (MMG). The MMG al-lows the automatic generation and modification of many different blended wing body aircraft variants. Furthermore, the MMG was able to automatically generate, for any aircraft variant, the specific models required by the project partners' analysis tools. Each analysis tools needed a different model coded in a different data format, which posed a number of pre-processing challenges. Besides, each tool had to be accessed via a web connection. The use of the MMG allowed the multidisciplinary analysis of 30 aircraft configuration in the time normally required to analyse just one.

These are the basic steps during the MMG operations:

1. The user (a designer in this case) submits an input file to the MMG, containing the values for the multitude of parameters used to define the aircraft configuration (e.g. wing span, position of engines and location of structural elements);

2. The user specifies the list of dedicated models (e.g. aerodynamic models and structural models) required for the various analysis tools;

3. The MMG (using the product model rules written in the ICAD language) automatically produces a blended wing body aircraft model, complete with internal structure and main systems;

4. On the basis of this unique model, the MMG generates a consistent set of dedicated models ready to be analysed by different analysis tools (the knowledge to pre-process the various models had previously been elicited from experts and captured in the product model);

5. The generated models are retrieved and analysed, and the results are used to prepare a new MMG input file (e.g. the wing span is modified or a different analysis model is specified);

6. The new input file is submitted to the MMG to generate an updated aircraft configuration with relative output models set.

## *3.3 How is a KBE application developed?*

**KEYWORDS:**
SYSTEM DEVELOPMENT, MOKA, PRODUCT MODEL

Most early KBE Systems were developed using methods devised by the particular people involved. There was no standardised approach and the success of an application was solely dependent on the abilities and experience of the development team. This obviously resulted in a significant risk when developing and maintaining KBE applications due to variations in people's abilities and experience.

A requirement emerged during the late 1990s for a standardised methodology. The leading KBE methodology to fulfil this requirement is called MOKA ("Methodology and tools Oriented to KBE Applications"). MOKA is particularly aimed at capturing and structuring design knowledge of complex mechanical products, in a way that facilitates the development of KBE applications.

MOKA [9] provides a methodology that aims to:

• Reduce the lead times and associated costs of developing KBE Systems;

• Provide a consistent way of developing and maintaining KBE Systems;

• Make use of a software tool to support the use of the methodology.

It does this by defining 6 phases for the development of a KBE System.

### 3.3.1 MOKA Project Phases

### 3.3.1.1 Phase 1: Identify

During this first phase, the key features of the project are identified, e.g. stakeholders, aims, requirements, source of knowledge and the target platform for the KBE system. The technical feasibility is assessed and the scope and purpose of the KBE system are defined.

### 3.3.1.2 Phase 2: Justify

During this phase, resource requirements and costs are estimated and project risks are assessed. A number of acceptance criteria are defined and a project plan is generated. A business case is prepared and approval is sought from management.

### 3.3.1.3 Phase 3: Capture

During this phase, knowledge is acquired from domain experts and modelled to create an "informal model", i.e. a structured representation made up of forms (called ICARE forms) and simple diagrammatic formats.

Within the informal model, the main knowledge objects are:

- Entities: These include Structural Entities (the components of the product being designed) and Functional Entities (the functions of the product and its sub-components);

- Constraints: The design requirements of the product and its sub-components;

- Activities: The tasks performed during the design process;

- Rules: The decision points in the design process that affect what tasks to perform;

- Illustrations: Specific examples that illustrate aspects of the product and design.

The informal model comprises two interlinked models:

- The **informal product model**: The structure of the product being designed, together with the constraints and functions associated with each component part;

- The **informal process model**: The activities that are performed by the design engineer including the rules involved at each stage.

3.3.1.4 Phase 4: Formalize

During this phase, a "formal model" is created. This uses a language called MML (MOKA Modelling Language), which is a graphical, object-oriented representation of engineering knowledge at one level of abstraction above application code.

The formal model takes the informal model as a fundamental input in order to represent the configuration of the KBE application itself. The formal model contains two distinct parts: The Product Model and the Design Process Model.

- The **formal product model** is used to specify the different features of the knowledge items in the informal model by using five "views": structure, function, behaviour, technology and geometric representation.

- The **design process model** is the process flow of the KBE system. This takes into account the constraints of the application, the operating system, the chosen programming language and the type of specialists involved in the process. This does not necessarily imply that all the knowledge involved in the manual process should be incorporated into the automated counterpart, as long as the result will be similar.

### 3.3.1.5 Phase 5: Package

During this phase, the programming code for the system is written, assembled, tested and debugged. To do this, the formal models created in the previous phase are translated (by hand or semi-automatically) into the appropriate structures and formats for the technology and programming language being used.

### 3.3.1.6 Phase 6: Activate

In this final phase, the KBE System is distributed to the end-user sites. The new system is introduced and the end-users are trained in its use.

## *3.4 What are the uses of KBE Systems?*

**KEYWORDS:**
PROCESS IMPROVEMENT, TASK AUTOMATION

KBE Systems are particularly useful when there is one or more of the following requirements:

- Reduction in the time required to design a particular product or component;

- Increase in the number of design concepts that are generated and analysed so as to create a better design solution;

- Improvement in the productivity of designers by increasing the time spent on creativity and innovation and decreasing the time spent on routine activities.

The most successful KBE applications involve the automation of routine designs that take up enormous amounts of times when performed manually. The pre-processing of analysis models is a very good example, particularly in the case of high fidelity analysis tools like finite elements methods (FEM) and computational fluid dynamics (CFD) codes.

The two case studies discussed in Section 3.2 provide good examples of typical situations where KBE is a winning technology. Some other examples are quoted by Stokes [9]:

- Textron Aerostructures reduced the lead-time of a tooling application by 73%;

- Jaguar cars reduced the design time for an inner bonnet from 8 weeks to 20 minutes;

- British Aerospace reduced the design time of a wing box from 8000 hours to 10 hours.

Vermeulen [7] quotes other examples of huge savings in time:

- Design of a windscreen wiper system from weeks to minutes;

- Optimisation of an airfoil shape from 2 months to 4 days;

- Design of a compressor from 10 days to 1 day.

However, KBE Systems are not the answer to all design situations. According to Stokes [9], you should not use KBE when:

- The design process cannot be clearly defined;

- The technology in the design process is constantly changing;

- The design process could just as well be modelled in a simple program;

- The knowledge for the desired application is not available;

- The organisation does not have the will, the money, and the resources to introduce a KBE system.

If the MOKA methodology is used to develop a KBE System (see Section 3.3) then the informal model can be used to generate a Knowledge Web (called a 'Knowledge Book') that provides a valuable source of information for engineers, and others, in the organisation. In this way, a project to create a KBE System has a number of other benefits, such as:

- Reduced risk of knowledge loss in areas where only one or two people hold vital knowledge in their heads;

- Increased efficiency in sharing knowledge in large, geographically-dispersed organisations;

- More awareness of engineering issues across the organisation;

- Improved training and induction of new design engineers;

- Better input to other initiatives, such as process re-engineering;

- A neutral knowledge representation to facilitate the migration of KBE applications from one KBE system to another.

## *3.5 What tools and technologies are available?*

**KEYWORDS:**
ICAD, AML, GDL, INTENT!, KNOWLEDGEWARE, KNOWLEDGE FUSION, PCPACK

### 3.5.1 KBE Systems and tools

A number of commercial tools and technologies are available to create KBE applications, or embed KBE functionality into a CAD System:

- **ICAD** (from KTI) came out in the early 1980's and was the first KBE system on the market. ICAD uses a programming language called IDL (ICAD Design Language), which is an object-oriented, declarative language based on LISP (List Processing Language). ICAD provides a proprietary CAD engine for surface modelling, and relies on the Parasolid system to handle solids. After KTI was bought by Dassault Systemes, ICAD has ceased to be supported, though many ICAD applications are still used by companies like Airbus and Boeing.

- **AML** (from Technosoft) is an object-oriented, knowledge-based engineering modelling framework. AML enables multidisciplinary modelling and integration of the entire product and process development cycle. According to Technosoft, no other commercial framework or development environment provides the full range of capabilities that AML includes out of the box.

- **INTENT!** is in some ways similar to ICAD. It is LISP-based and was developed by people who worked on the development of ICAD. INTENT! uses AutoCAD as its geometry engine.

- **GDL** (from Genworks) is a new generation KBE system that combines the power and flexibility of the older ICAD system with new web technology. It is available for many different platforms such as Windows, Linux and Mac. Its programming language is based on the standard ANSI Common LISP. It allows the manipulation of very simple geometry primitives, and optionally provides full integration to the NURBs Surfaces and Solids modelling kernel from Solid Modeling Solutions Inc.

- **KnowledgeWare** (from Dassault Systemes) is a set of applications available to extend the native functionalities of the CATIA V5 CAD system in terms of design automation and rules capturing. Knowledgeware offers the possibility to define product templates so that automated parametric design is facilitated. Other tools are provided to organise and manipulate parameters, create flexible rules and specification checks. Apart from Knowledgeware, CATIA V5 also offers designers the possibility to write pieces of Visual Basic to further extend the design automation capability.

- **Knowledge Fusion** is a KBE system that is part of Unigraphics, one the leading CAD systems on the market. Unigraphics uses INTENT! as the modelling language and Parasolids as the geometry engine. The user can build interfaces for different geometric parts. Databases are easy to access by using an Open Database Connectivity (ODBC) method.

3.5.2 Augmented CAD Systems

As described in the previous section, several of the major CAD companies now market products that offer KBE func-tionality. These hybrid systems may sound like a promising idea. Indeed, they do contribute to the level of automation in the world of Computer Aided Design, where the amount of "donkeywork" is still very high. They are also significantly contributing to the diffusion of KBE technology, which is finally reaching other customers than the usual aircraft and automotive giants.

However, these systems are born and developed as extensions to the main CAD capabilities, rather than being actual KBE systems. Hence, according to Cooper and La Rocca [10], they still have some major deficiencies:

- KBE applications generated with these augmented CAD systems run quite inefficiently. They are typically some orders of magnitude slower than real compiled KBE applications;

- In general, they do not provide full generative modelling, which is a key feature of a true KBE system. In fact, they do not allow exploitation of all the CAD functionalities via their KBE programming language, and require a lot of manual operation (i.e. many menu clicks);

- They are generally supported with a rather narrow choice of computer and operating system platforms (e.g. Linux is typically not an option);

- They are not typically web-friendly;

Taken as a whole, these criteria help to clarify the distinct and complementary roles of true KBE on one hand, and CAD on the other.

3.5.3 Support tools for KBE Development

The MOKA methodology is supported by PCPACK (see Section 2.5.1). PCPACK can be used to create all of the in-formal models using a combination of the Annotation tool (to create ICARE forms) and the Ladder and Diagram tools to create the diagrammatic representations.

The Diagram tool can also be used to create all of the formal MOKA models. The formal product model is created using 5 different diagram templates that represent the 5 views of the formal product model. The formal process model is created using activity diagrams. These models are stored in an XML format, which can be translated (e.g. using XSL stylesheets) into code for use when developing a KBE application. Though the feasibility of this translation has been demonstrated, more work is required to produce a robust, generic and straightforward way of passing from knowledge elicitation to an automatically-generated KBE application.

## *3.6 What are the issues of KBE Systems?*

**KEYWORDS:**
ISSUES, CAD, MOKA

The first KBE Systems were written in the 1980s. Since then, there have been some impressive systems that have saved significant time and cost (see Section 3.4).

However, KBE Systems have not taken off as a widespread technology, as compared to CAD technology. Why is this? Often KBE has been used just as a final weapon to meet the deadlines and deliverables of critically complex projects, rather than becoming a technology of daily use.

The programming approach of true KBE Systems provides the most flexible and powerful way of generating dedicated design solutions. However, it requires extensive training and highly-educated developers. Although there have been many good pilot projects, it is probable that the large in-vestments needed for the development of KBE Systems has been a deciding factor for managers.

To overcome this problem, those interested in KBE have been working to reduce the resources and risks associated with the development of a KBE application and to maximise the benefit of a single project. Hence, re-use of generic knowledge has been a major aim, as has the development of methods and tools to create a more structured, efficient and effective development process.

The MOKA methodology is a major landmark in this endeavour, however its uptake has not been as great as might have been envisaged. Although it contains many good ideas, it still lacks some maturity. MOKA users (such as Airbus and Fokker) have been refining and customising it for their particular context and objectives. Useful additions have included the use of special elicitation techniques to interview engineering specialists and software modelling tools such as PCPACK. It remains the case, however, that the biggest impact of MOKA has probably been on Knowledge Management, i.e. creating web-based material to share engineering knowledge around an organisation. However it is likely that there will be a dramatic impact as soon as the challenge is met of the automatic generation of a functioning KBE application directly from a knowledge base.

The relationship between KBE and CAD Systems is an important one. In the words of Cooper and La Rocca [10]:

> Historically there has been ambiguity and controversy regarding the role of KBE with respect to its interaction with traditional CAD systems. Whatever the past confusion of roles, the reality of the engineering and design workplace today is that they are largely CAD-based. Drumbeats such as "CAD the Master" were widely popular in the 1990's, and to a large extent, still are. Consequently, to be successful in the current marketplace, state-of-the-art KBE technology must complement the existing CAD systems, just as it must complement existing Database systems (e.g. Oracle, MySQL), delivery mechanisms (e.g. web servers and web browsers), and typesetting formats (e.g. PDF).

KBE is a young discipline, hence technologies and methodologies are still being developed and refined. However, the drivers for this development are strong given the huge potential that KBE Systems have to revolutionise the product development process.

## *3.7 Where can I get more information?*

See references 6-10 in the Bibliography (Section 8).

Other sources of information:

Chapman, C.B. and Pinfold, M. (1999). Design engineering - a need to rethink the solution using knowledge based engineering. *Knowledge-based Systems*, Vol. 12, pp.257-267.

Cooper, S., Fan, I. and Li, G. (2001). Achieving Competitive Advantage Through Knowledge Based Engineering - A best practice guide, *White Paper for the Department of Trade and Industry*. University of Cranfield, UK.

## 4. KNOWLEDGE WEBS

*4.1 What is a Knowledge Web?*

**KEYWORDS:**
KNOWLEDGE WEB

A Knowledge Web is a website that is generated automatically from a knowledge base. The aim of a Knowledge Web is to present the information contained in the knowledge base to a human user in a clear and navigable way.

A Knowledge Web is similar to an ordinary website but differs in a number of important respects:

1. It is more structured and meaningful;

2. It contains knowledge from expert practitioners;

3. It contains fully validated knowledge, often combined from a number of experts;

4. It provides the user with different ways of viewing the knowledge base.

Let us examine some of these ideas.

4.1.1 Structure and Meaning

As described in Section 1.4, the basic elements of a knowledge base are concepts, attributes, values and relations. A Knowledge Web retains the essential structure of the knowledge base and so presents the relevant expertise in a clear and meaningful way.

For example, each concept has its own web page (called an annotation page) which is a frame-like representation showing its properties (attributes and values) and the relations between it and other concepts. Each class of objects has the same struc-ture, i.e. has the same headings for each of the rows. Some examples are shown in Table 2.

Table 2. Examples of headings used in annotation pages

| Headings for a Task | Headings for a Role | Headings for a Document |
|---|---|---|
| Description | Description | Description |
| Inputs | Held by | Produced by |
| Produces | Performs | Used by |
| Triggers | Uses | Format |
| Followed by | Works for | Resource for |
| Preceded by | Located in | Located in |
| Performed by | Has expertise | Owned by |

The hyperlinks between pages have a meaning that is shown by the headings in the left-hand column of the table, i.e. hyperlinks represent relationships.

### 4.1.2 Validated Expertise

As described in Section 4.3, the underlying knowledge base is developed in a methodical and systematic manner. This involves two important elements: (i) special techniques to elicit deep, tacit knowledge; (ii) an on-going process of validation to ensure the captured knowledge is correct, complete, consistent and relevant.

These two features of the development process ensure the Knowledge Web reflects the content and structure of the knowledge in the heads of domain experts. In this way, the Knowledge Web provides a rich environment for learning about the domain (see Section 4.6.2). It also distinguishes a Knowledge Web from other web resources (such as most web pages, Wikis and Blogs) in that every part is guaranteed to hold validated expertise that is relevant to the requirements of the organisation.

### 4.1.3 Multiple Viewpoints

Most web sites describe information using text with occasional images. A Knowledge Web combines textual descriptions in annotation pages with various graphical representations such as:

- Trees (hierarchies of concepts)
- Diagrams (networks of concepts)
- Matrices (the relations that exist between two sets of concepts)

When these formats are expressed in special web languages, such as SVG or XAML (see Section 4.5.3), then automatic hotlinks can be generated from nodes to annotation pages making navigation much easier and more fruitful.

In addition to predefined trees, diagrams and matrices, special ways of displaying the knowledge can be created from user-defined parameters and selections. This is achieved using the latest web technologies such as XML and XSL (see Sections 4.5.1 and 4.5.2). These technologies allow the user to choose personalised ways of viewing and navigating the knowledge base.

## *4.2 What is the structure of a Knowledge Web?*

**KEYWORDS:**
ANNOTATION PAGES, TREES, PROCESS MAPS, CONCEPT MAPS

Different people want to use a Knowledge Web in different ways for different purposes. For example, some people want to go directly to a small piece of information, whereas others want to move through the information in great detail learning as much as they can. These different uses are discussed later in Section 4.4. As far as the structure is concerned, it must be designed in a way that provides users with:

- Different ways to find what is wanted;
- Different ways to view the contents of the knowledge base.

Let us look at each of these.

4.2.1 Search and Navigation Options

To provide the user with a selection of ways to find what is wanted, the following features can be used:

- A 'Search' facility – either a simple text-based (key word) search or a semantic-based search (making use of techniques described in Chapter 6);
- An 'A-Z List' of the pages and diagrams;
- A 'Glossary' of key terms with hyperlinks to annotation pages;
- Hot links on diagrams (linking to annotation pages);
- A 'Browser Tree' that shows the overall taxonomy or other user-defined trees (e.g. a task decomposition or product decomposition);
- Hyperlinks in each page to other relevant pages or to external files (e.g. pdf documents, spreadsheets, video files).

4.2.2 Viewing Options

A number of representations can be used to provide the user with different ways of viewing the contents of the knowledge base. Such representations as annotation pages, trees, diagrams, matrices and dynamically created pages can be incorporated into a Knowledge Web. Let us take a look at these.

### 4.2.2.1 Annotation pages

Each concept in the knowledge base has an annotation page that shows: (i) the attributes and values of the concept, (ii) the relations of the concept to other concepts. Pages for concepts of the same type (class) are structured in the same way (as was described in Section 4.1.1). This ensures that information is presented in a clear and accessible way. The frame-like structure means that fields (rows) can be populated automatically with the contents of the knowledge base (using annotation templates). Hyperlinks to other related annotation pages provide for easy navigation. An example of an annotation page is shown in Figure 3.

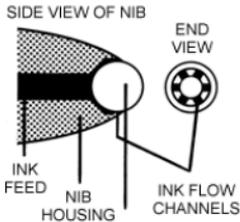

Figure 3. Part of the annotation page for a task

### 4.2.2.2 Trees and Diagrams

Trees and diagrams are often used when acquiring and validating knowledge from domain experts (see Section 4.3.2). These formats show the concepts in the knowledge base as nodes and their inter-relationships as links or arrows. Common examples are:

- **Concept tree**: This is a hierarchy where every link has the 'is a' relation. Hence, it shows the classes to which every concept belongs. This form of knowledge is called a taxonomy and is a key aspect of a knowledge base.

- **Composition tree**: This is a hierarchy where every link has the 'has part' relation. This is used to show the components and sub-components of a concept such as a complex product, a document or an organisation.

- **Process tree**: This is a special form of composition tree in which all the nodes are tasks. Hence, it shows how a complex task is composed of sub-tasks and sub-sub-tasks, etc.

- **Concept map**: This is a diagram that shows a variety of concepts connected by a mixture of different relations. Concept maps can come in different varieties, such as hierarchical concept maps and those that restrict the concepts and relations that are shown.

- **Process map**: This type of diagram shows the way a task (process, activity) is performed. The main elements on a process map are the sub-tasks of the task that is being described. These sub-tasks are placed on the map in the order in which they are performed. Links between tasks represent the 'followed by' rela-tion. Other concepts can be included on a process map to show further information, such as the resources, products and roles involved in a task.

Trees and diagrams are incorporated into a Knowledge Web by means of technologies such as SVG or XAML (see Section 4.5.3). This provides images that are interactive, with automatic hotlinks between trees/diagrams and annotation pages. Once again this provides a good environment for the user to find what is required and to navigate around areas that are semantically close.

### 4.2.2.3 Dynamically-created pages

Dynamically-created pages are another way for users to find useful information. This can be achieved by presenting the user with a form, such as a number of pull-down boxes, so that he/she can select particular settings. For example, the options might be the person's role, a project phase and a topic of interest. These settings provide variables that can be used to select or customise a stylesheet that will show the most relevant contents of the knowledge base in the most relevant way. This can be achieved using the technologies described in Sections 4.5.1 and 4.5.2.

### 4.2.3 Overall Knowledge Web Structure

The previous sections have described a number of elements that can form part of a Knowledge Web. Figure 4 shows how these different elements can be linked together to form a complete Knowledge Web. The dotted lines in Figure 4 represent hyperlinks.

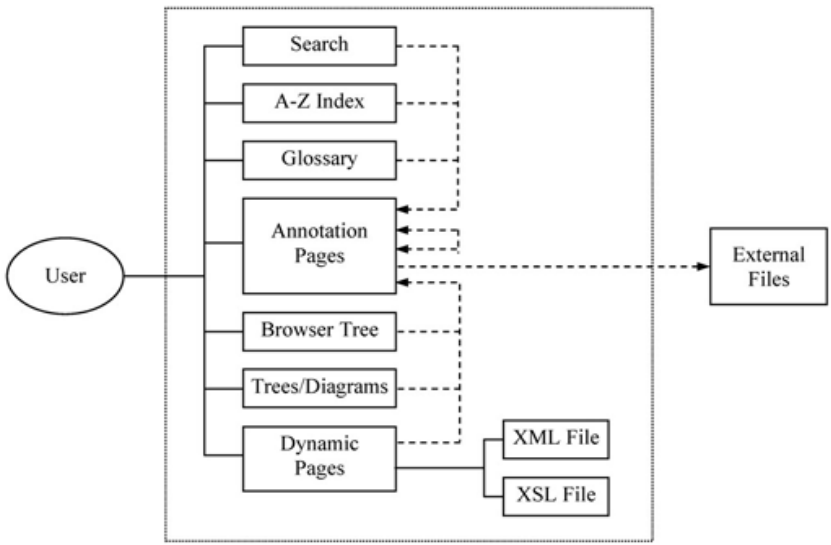

Figure 4. Components and Structure of a Knowledge Web

## *4.3 How is a Knowledge Web developed?*

**KEYWORDS:**

SYSTEM DEVELOPMENT, 47-STEP PROCEDURE, SCOPING, KNOWLEDGE ACQUISITION, INTERVIEW TECHNIQUES, KNOWLEDGE MODELS, META-MODEL, VALIDATION, TACIT KNOWLEDGE

In the book Knowledge Acquisition in Practice [2], I describe a 47-step procedure for developing a knowledge base and the resulting end-product. This procedure can be used to create Knowledge Webs (and other knowledge resources) for any domain, situation or organisation. The procedure is split into four phases, as described below.

4.3.1 Phase 1: Start, Scope and Plan the Project

The first steps involve identifying a project idea and documenting it as a project proposal. After discussion and modifi-cation of the proposal, the initial phase of knowledge capture begins in order to scope the project, i.e. define which specific areas of knowledge will be acquired.

The sources of knowledge needed for the project are identified, and a project plan is created.

### 4.3.2 Phase 2: Initial Capture and Modelling

The Knowledge Engineer begins this phase by learning the basics of the domain from documents or informal conversations with domain experts then uses semi-structured interviews with domain experts. The captured knowledge is analysed to identify the key concepts. A taxonomy, in the form of a concept tree, is created and validated with domain experts.

Further rounds of interviews, analysis and knowledge modelling are performed. During this a **meta-model** is defined that shows how the knowledge will be represented and structured. The relationships between concepts and properties (attributes and values) of concepts are modelled using the appropriate knowledge models (e.g. trees, diagrams, matrices). These are used in an on-going iterative process to validate (check) the knowledge with domain experts and add new knowledge. A first-pass knowledge base is now in place, which may be all that is required for some small projects.

### 4.3.3 Phase 3: Detailed Capture and Modelling

If required, further interviews and modelling activities are used to capture more detailed knowledge. A prototype Knowledge Web is created and used to carry out an assessment exercise with a representative sample of end-users.

If necessary, the acquisition of very detailed (deep, tacit) knowledge now takes place using a suitable set of specialised interview techniques, such as concept sorting, repertory grid and protocol analysis. If required, cross-validation takes place, i.e. other experts check that the knowledge base developed with the main experts is correct, complete, relevant and best practice.

### 4.3.4 Phase 4: Share the Store Knowledge

The final phase starts by defining and creating the format of the Knowledge Web. A provisional Knowledge Web is created and given a full assessment by end-users. After improvements have been made, the finalised Knowledge Web is released for use in the organisation. The Knowledge Web is publicised to the organisation so that all potential end-users know what it is, where it is and how it can be used.

After the Knowledge Web has been used for some time, its impact on the organisation is assessed and documented. Finally, a complete project review takes place to learn lessons and make suggestions that can be used to improve the methodology and the support systems.

## *4.4 What are the uses of Knowledge Webs?*

**KEYWORDS:**
KNOWLEDGE SHARING, CORPORATE MEMORY

The act of creating and then using a Knowledge Web provides a variety of uses and benefits to an organisation:

- Presenting expertise to the people that need it;

- Helping inexperienced people to perform more like an expert;

- Teaching inexperienced people more about a domain;

- Accelerating the learning process;

- Helping to build the corporate memory (e.g. archiving

knowledge for future generations);

- Establishing best practices;

- Helping people solve specific problems;

- Reducing the risk of knowledge being lost from the organisation.

Let us examine some of these in more detail.

### 4.4.1 Teaching and Learning

A Knowledge Web can be used to teach people who are just starting in an area and accelerate their progress up the learning curve.

The teaching aspect can be accelerated further if a starter in the organisation performs all the activities required to build the Knowledge Web, i.e. he or she is the knowledge engineer on the project (after a suitable period of training in knowledge acquisition). In this way, not only is a useful resource created, but the new starter can reach a level of proficiency much more quickly than normal (as much as 4 times faster).

### 4.4.2 Dissemination

A Knowledge Web can be used to spread knowledge across the functional boundaries of an organisation, such as from design to manufacturing (and vice versa), or from technical people to financial people (and vice versa). In this way, an expert in one area can benefit from expertise in another area.

### 4.4.3 Corporate Memory

A Knowledge Web can be used to archive knowledge for future generations. For example, it can store the reasons behind decisions that are made during the development of a new product. Some complex products such as military and aerospace products have an active life of many decades so it is vital that knowledge is passed down the generations in a format that is useable over a long span of time, i.e. is easily accessed, searched and understood.

A Knowledge Web can be used to collate lessons that have been learned on past projects so that mistakes are not repeated and a new set of people can benefit from the past experience of other people.

### 4.4.4 Reducing Risks

A Knowledge Web can be used to reduce the risks involved in losing access to people who have very specific knowledge. As such, it is useful to create a Knowledge Web when people are close to retirement or when there are only one or two experts in a specific area.

### 4.4.5 Problem Solving

A Knowledge Web is useful when there is a problem that requires information and knowledge to be gathered and collated from several sources (people, documentation and databases) and then presented to those who can solve the problem. For example, Knowledge Webs have been created to address specific problems in design and manufacturing areas that have led to significant savings in time and cost.

## *4.5 What tools and technologies are available?*

**KEYWORDS:**
XML, XSL, SVG, XAML, PCPACK

A Knowledge Web can be delivered over the Internet, on an Intranet system (the most likely) or on a CD. Whatever the case, it will be viewed using an Internet browser (such as Internet Explorer or Mozilla Firefox). Hence it must use standard Web formats. Annotation pages can use the most common format, HTML (Hypertext Mark-up Language). Active features, such as search, can be implemented using Web programming languages such as JavaScript.

To provide users with the most relevant and useful information, the Knowledge Web can incorporate the latest Web technologies, such as XML, XSL, SVG and XAML. Let us take a look at these.

### 4.5.1 XML

XML means "Extensible Mark-up Language". It is a Web standard for storing meaningful information about a group of concepts or about the contents of a document. It does this by surrounding the different elements in a text file (words, phrases, sentences, etc.) with tags. A tag is a word or phrase in angled brackets.

For example, suppose I have the following sentence in a text file. "Manchester United is a football team in England". I can add tags to parts of the sentence to show more information:

"<football club>Manchester United</football club> is a <sport>football</sport> team in <country>England </country>".

HTML (the standard web page format) also uses tags but these describe how a piece of text will be displayed, not information about it. This difference between HTML and XML is highlighted in Table 3.

Table 3. Differences between HTML and XML

| | HTML | XML |
|---|---|---|
| **Used for** | Web pages | Internet and Intranet databases<br>Knowledge Bases<br>Translation formats |
| **Representation** | Tags represent how text should appear on a web page | Tags represent the meaning of pieces of text |
| **Example tags** | \<p\> for paragraph<br>\<b\> for bold<br>\<tr\> for table row | \<statement\> for statement<br>\<city\> for city<br>\<rule\> for rule |
| **Examples** | \<p\>\<i\>The meaning of life\</i\> is a \<b\>meaningless\</b\> sentence.\</p\> | \<statement\>The meaning of \<concept\>life\</concept\> is a \<value\>meaningless\</value\> sentence.\</statement\> |

As shown in the table above, both HTML and XML can have embedded tags (i.e. tags within tags). As with HTML tags, XML tags can also include attributes to show the properties of the text being tagged. For example:

\<oil viscosity="75" cost="high" \>Oil 344\</oil\>

When designing the format of the XML, a choice of what tags and attributes to use must be made. To illustrate the different options, the following 3 XML statements have the same semantics (i.e. meaning) but it is shown in different ways:

- \<book\>\<title\>The Magus\</title\>\<author\>John Fowles\</author\>\</book\>

- \<book author="John Fowles"\>The Magus\</book\>

- \<book title="The Magus" author="John Fowles"/\>

### 4.5.2 XSL

XML files are great for storing extra information about the contents of a web site, but they are no good unless we can do something with them. In a Knowledge Web, we want the users to be able to view those parts of the XML knowledge base that will be of most use. People will have different requirements so they need different ways of viewing the knowledge base. To do this, we use XSL stylesheets.

XSL means "Extensible Stylesheet Language". It is a way of transforming and displaying XML files. Different XSL stylesheets, when applied to the same XML file, show the contents of the file in different ways.

Each XSL stylesheet does two things: (i) Searches for items in the XML file that will be displayed; (ii) Defines the way in which the items will be displayed.

A simple example of this is shown in Table 4. In the left-hand column is a piece of XML describing some countries. In the middle column is a piece of XSL that can be used to create the text shown in the right-hand column.

Table 4. Examples of XML, XSL and the resulting output

| XML | XSL | OUTPUT |
|-----|-----|--------|
| <country> | <body> | **Countries** |
|  <name>England</name> | <h1>Countries</h1> | England |
|  <pop>51 million</pop> | <xsl:for-each select="//country/name"> | Italy |
|  <capital>London</capital> |  <xsl:value-of select="."/> | China |
| <cont>Europe</cont> |  <br/> | |
| </country> | </xsl:for-each> | |
| <country> | | **Capital Cities** |
|  <name>Italy</name> | | London |
|  <pop>58 million</pop> | <h1>Capital Cities</h1> | Rome |
|  <capital>Rome</capital> | <xsl:for-each select="//country/capital"> | Beijing |
|  <cont>Europe</cont> |  <xsl:value-of select="."/> | |
| </country> |  <br/> | |
| <name>China</name> | </xsl:for-each> | |
|  <pop>1322 million</pop> | </body> | |
|  <capital>Beijing</capital> | | |
|  <cont>Asia</cont> | | |
| </country> | | |

Using the same XML as shown in the table above, a different XSL stylesheet could be written to create the matrix shown in Figure 5.

|  | Less than 30 million people | 30-80 million people | More than 80 million people |
|---|---|---|---|
| **Europe** |  | • England<br>• Italy |  |
| **Asia** |  |  | • China |

Figure 5. Example of a matrix created using an XSL stylesheet

XSL files can take in user-defined variables, e.g. with pull-down boxes on a web form. In this way, information can be presented that is personalised to the user's requirements. For instance, a form plus XSL stylesheet plus XML file can do the following:

- The user wants to see a matrix like the one shown in Figure 5, but would like to able to select different attributes for the columns and rows;

- The user wants to see a list of those HTML pages on a Knowledge Web that are most relevant to his/her situation. This can be achieved by storing information in the XML about the relevance of each page to different users, e.g. as attributes representing a user's role, project phase and topic area;

- The user is a programmer who would like the Knowledge Web to generate software code automatically from the knowledge base. This is achievable by creating an XSL stylesheet that shows the XML contents using the syntax of a programming language.

### 4.5.3 Graphical Mark-up Languages

In Section 4.2.2.2 we saw that trees and diagrams are useful ways of showing information. Graphical mark-up languages allow these to be shown using web formats. The two main languages are SVG and XAML.

SVG means "Scalable Vector Graphics". It is a Web standard for two-dimensional graphics written in a specialised form of XML. A web browser renders an SVG file to display shapes and images. These can be animated and made interactive.

XAML is also an XML-based file format for describing 2D graphics. Unlike SVG, XAML does not include a programming API for graphical applications, but it does support things like 3D and controls.

SVG and XAML allow the nodes on trees and diagrams to be automatically hot-linked to annotation pages, and allow functions such as search, pan and zoom.

### 4.5.4 PCPACK Publishing Tool

As described in Section 2.5.1, PCPACK is a suite of tools that helps a knowledge engineer to capture knowledge and create a knowledge base. PCPACK includes a tool called the Publishing Tool that transforms a knowledge base into a Knowledge Web. To do this it uses all of the technologies described above. Let us see how this happens.

The Publishing Tool creates a Knowledge Web by applying the contents of the knowledge base to a template. This publication template includes JavaScript and XSL files to transform the contents of the knowledge base into the views and functions required by the web users (e.g. browser tree, search facility, A-Z index and glossary). The template also includes HTML files (such as the banner) and CSS file to give the Knowledge Web the right look (e.g. colours, font, logos).

Once a template has been created for an organisation, it will be used for all its Knowledge Webs (with minor modifications, such as project-specific XSL files). In this way, a knowledge engineer can publish (i.e. transform) a PCPACK knowledge base to create a Knowledge Web in a matter of minutes and requiring no Web skills. This not only reduces the time and cost of creating Knowledge Webs, it also means prototype webs can be created at any point in the project for assessment with end-users, thus providing vital feedback and a better end-product.

## *4.6 What are the issues of Knowledge Webs?*

**KEYWORDS:**
ISSUES, USABILITY, BUSINESS BENEFITS

The key issue for a Knowledge Web is this: How can highly-structured information be presented in ways that maximise its usefulness and usability for different users in different situations with different requirements?

Knowledge Engineers address this issue in three main ways.

1. Provide a structure to the web site;

2. Provide the user with various ways of finding and viewing the knowledge;

3. Develop the Knowledge Web in a way that involves end-users and domain experts.

Let us take a look at these three ideas.

### 4.6.1 Structure

When designing a Knowledge Web, it is important to ac-

knowledge the problems that users might have. For example, a user might:

- Miss some vital pieces of knowledge;

- Spend too long searching for what is required;

- Read irrelevant information but think it is relevant;

- Gain a fragmented and incoherent understanding of the knowledge.

The usual format for a web site is that information is presented non-sequentially, i.e. there is no beginning and no end (as in a book). The advantage of this is obvious – users can navigate around following links they find most interesting. However, there are disadvantages to this: the user might not take a good route through the material, might get lost, and might miss important areas.

It is important, therefore, that a Knowledge Web has a structure, so that the user knows where they are, knows where they have been and can see different areas they might want to visit. Since the web is being constructed from a knowledge base, structures are already present. Thus the user should be provided with either: (i) a single main structure, (ii) a selection of structures to use; or (iii) both a main structure and an opportunity to change this.
Here are some options:

- **Taxonomy**: The main structure for many Knowledge Webs is a taxonomy, i.e. a classification tree. It is usual to have this as a browser tree on the left of the screen so that users can expand different branches that represent different types of concepts, i.e. different types of pages.

- **Decomposition Tree**: Some Knowledge Webs have as their main structure a decomposition of one of the

main things in the domain. This can be a product (e.g. an engine, a document), or a process (e.g. design the engine, write the document). A MOKA Knowledge Web (see Section 3.4) provides both of these – a product view and a process view.

- **Process Flow**: Alongside hierarchical structures, many knowledge webs have process maps (see Section 4.2.2.2) that allow the user to navigate to lower levels (i.e. see diagrams of more detailed activities) or to higher levels (i.e. to see more general activities).

- **Special Navigation Tree**: It is sometimes useful to construct a special tree that gives the user all the main pages they might want to see, in a clear hierarchy, or that takes a sub-set of another of the trees, so the user is not overwhelmed with too much information.

- **Overview Map**: Just as a special tree can be created for navigation purposes, so can a special diagram, on which each node represents a key page or diagram, to which the node is hot-linked.

## 4.6.2 Multiple views and functions

In Section 4.2, we saw the structure of a typical Knowledge Web. Several options were described that provide the user with a choice of different ways to find and view the knowledge. Why have such a big choice? The answer can be seen if we consider the user's requirements along two dimensions:

- **Breadth of knowledge**: Does a user want to view a narrow area of the domain or look across the whole domain?

- **Depth of knowledge**: Does a user want a basic level of understanding or a very detailed (deep) level of understanding?

Table 5 shows these 2 dimensions, and the resulting 4 types of requirement as segments. Each segment summarises the requirements of the user and the approach taken in the Knowledge Web to satisfy the requirements.

Table 5. Web user's requirements shown as a 2x2 matrix

|  | Narrow piece of knowledge | Broad piece of knowledge |
|---|---|---|
| **Basic level of knowledge** | **Requirement**: The user needs to find a small amount of information.<br><br>**Approach**: The structure helps the user find the information using the browser tree, A-Z list, search facility or dynamic (XSL-generated) pages. | **Requirement**: The user needs to gain a general feeling for the whole domain in a short amount of time.<br><br>**Approach**: Use of a hierarchical structure allows the user to navigate across the upper levels only. |
| **Deep level of knowledge** | **Requirement**: The user needs a specific piece of detailed knowledge<br><br>**Approach**: The browser tree, A-Z list and search facility are useful, but this is where dynamic (XSL-generated) pages are of particular use. | **Requirement**: The user needs to learn everything in the Knowledge Web.<br><br>**Approach**: The structure provides a map of where to go. A special sequential structure can be added to provide a beginning, middle and end. |

Although it is generally a good idea to provide the user with multiple routes for finding and viewing the knowledge, there is a danger that the user is overwhelmed with too many options and operational complexity. This brings us to the third of the important factors – a good development process.

4.6.3 Development Process

It is no good spending weeks creating a Knowledge Web full of expertise without it also satisfying three important criteria: (i) Will people be able to access and use the Knowledge Web? (ii) Will the Knowledge Web be useful for people? (iii) Will people want to use the Knowledge Web? To address these questions, a number of key activities need to be performed during the development stages.

1. During the project definition and scoping phase it is important to (i) create a good project proposal with input from all key people; (ii) define a project scope by involving domain experts and end-users; (iii) create a good project schedule that maximises the resources available.

2. During the acquisition and modelling phases it is important to (i) use effective methods for eliciting, modelling and validating knowledge; (ii) use special methods for eliciting deep, tacit knowledge; (iii) Ensure that all of the k-base contents are correct, complete and best practice; (iv) review the project aims so they can be re-aligned for maximum benefit.

3. During the final phases it is important to: (i) create and assess a prototype Knowledge Web with end-users; (ii) assess the final Knowledge Web with end-users before making final modifications prior to release; (iii) release the Knowledge Web in the right way and provide information and training; (iv) publicise the Knowledge Web to all potential end-users; (v) post release, assess the impact of the Knowledge Web and make any necessary modifications.

## 4.7 Where can I get more information?

See reference 2 in the Bibliography (Section 8).

Other sources of information:

Milton, N., Shadbolt, N., Cottam, H. and Hammersley, M. (1999). Towards a Knowledge Technology for Knowledge Management. *International Journal of Human-Computer Studies*, volume 51, pp. 615-641.

Dykes, L. and Tittel, E. (2005). *XML for Dummies*, 4th edition. Hoboken, NJ: Wiley.

# 5. ONTOLOGIES

## 5.1 What is an Ontology?

**KEYWORDS:**
ONTOLOGY

An ontology can be thought of in two ways:

- A specification of how the knowledge in a domain can be modelled (represented, described, structured);

- A type of file embedded in a knowledge-rich IT system that provides important information for other parts of the system.

Hence, people use words such as framework, schema, vocabulary, conceptualisation, meta-model and skeleton when describing ontologies.

Each ontology focuses on a particular domain (topic, discipline, subject area). For example, an ontology of medicine might include such things as:

- The main types of diseases and their sub-types;

- Different diagnostic techniques;

- The attributes used to describe patients, such as blood pressure, gender and age;

- Types of doctors, such as dermatologist and radiologist;

- Types of equipment used by doctors;

- Types of treatment, such as antibiotics and radiation therapy;

- The relationships between these things, such as which

diseases should be treated by which doctors using which diagnostic techniques and which equipment;

- Rules or axioms, such as male patients cannot be pregnant, and babies cannot be alcoholics.

Unlike the other knowledge technologies in this book, ontologies do not carry out activities that directly interact with a user. They are an enabling technology; operating out of sight, behind the scenes. As such, it can be hard for people unfamiliar with computer science and knowledge technologies to appreciate their value. Here are some examples that illustrate how they form an integral part of many successful knowledge technologies:

- An ontology can be used when developing a Knowledge Based System to define what knowledge to capture and how to represent it. For instance, in the CommonKADS methodology ontologies are combined with Problem Solving Models to provide a library of re-usable knowledge models [3];

- An ontology derived from the MOKA methodology [9] can be used as the starting point for the development of a KBE System;

- An ontology can be the starting point for a knowledge base that is used to produce a Knowledge Web. For example, an ontology template is used as the starting point for each knowledge base in PCPACK [4];

- Ontologies are an essential element of the Semantic Technologies described in Chapter 6. They are embedded within software systems, providing information and structure to other parts of the system, such as databases, inference engines, web services, information agents, and user interface tools.

As you may have realised, an ontology is very similar to a knowledge base (see Section 1.4). The key differences are:

1. An ontology often has a more formal structure than a knowledge base. This is because ontologies are used by computer systems, and computer systems need to be given a lot of information that people take for granted, and in a format a computer can process;

2. An ontology is usually a general model of knowledge and contains no specific information. For example, a knowledge base might hold information on actual patients, their attributes and the treatments they are having, but an ontology would not include such case-specific information.

Some other ways of thinking about an ontology are:

- As a taxonomy (tree of classes) of the objects in a domain, plus other relationships between the classes (e.g. causes, part of), plus attributes and values for each of the classes;

- As a vocabulary that has a grammar and structure;

- As a very rich form of database schema that defines a commonly held view of a domain. Fensel [11] has observed, that an ontology is a database schema for the 21st century that: (i) uses a richer language than a database schema, (ii) is not in a tabular format, (iii) uses a shared terminology and conceptualisation from a number of domain experts;

- As a "formal, explicit specification of a shared conceptualization" [12].

The last two bullet points convey an important aspect of an ontology: that it is usually a shared vision from a number of domain experts, using a vocabulary that all agree upon.

## 5.2 What is the structure of an Ontology?

The structure of an ontology will depend on whether it is frame-based or logic-based (see Section 1.4.2). To simplify things, and not delve into the complexities of these two approaches, I will provide a basic overview that mixes the two approaches (apologies to purists for this!).

To give you an idea of the structure and contents of an ontology, let us look at a specific example.

In this example, the ontology is being used to combine knowledge from different sources, so that a software application (e.g. a web service) can do something with the integrated picture.

Suppose there are 2 websites that both describe cars (automobiles). One of them has been written by a UK company for a UK audience; the other by an American company for an American audience. Here is a brief description of their contents and structure:

- The UK website uses terms like "bonnet", "petrol pump" and "bumpers" for parts of a car. It has categorises such as "sports car", "saloon car" and "MPVs". Underlying the website is a knowledge base that represents the manufacturer of a particular model as objects and uses the "makes" relation to associate a model with a manufacturer. The knowledge base describes the "engine capacity" and "maximum speed" of a car with attributes and ordinal (numerical) values. Figures for "maximum speed" are given in km/hr.

- The US website uses terms like "hood", "gas pump" and "fenders" for parts of an automobile. It categorises them into classes based on where the manufacturer is located, e.g. "US", "Japanese", "European". Underlying the website is a database storing data on manufacturers and models. The "engine size" and "max speed" for each type of automobile are described using text on a web page. Figures for "max speed" are given in mph.

Suppose you are given the task of connecting these two information resources together so they can share information. You have to write a software application to do this. As you may have guessed, the best way to go about this is to develop an ontology about cars. This ontology needs to be able to deal with:

1. Different names for the same thing (e.g. 'bonnet' is a synonym of 'hood');

2. Different taxonomies (ways of classifying things);

3. Different ways of representing things (e.g. as relations in a knowledge base, as a table in a database, as a piece of html in a web page);

4. Different units for numerical values (e.g. km/hr and mph).

So to provide a way of specifying the language and structure to be used across these different information resources, the ontology must define:

- The names of objects (including any synonyms);

- A general taxonomy that all experts are happy to use;

- A way of representing relationships and/or attributes and values;

- Any rules, constraints or axioms.

One way to show some of the ontology is by using frames. A frame is a table that shows the attributes and values for a class of objects in the ontology. 3 examples of frames (for 'car', 'engine' and 'manufacturer') are shown in Figure 6.

| car | engine | manufacturer |
|---|---|---|
| Class: *vehicle* | Class: *car component* | Class: *organisation* |
| Number passengers: *Ordinal* | Type: *{piston, wankel, other}* | Nationality: *Categorical* |
| Max speed (mph): *Number* | Capacity: *Number* | Prestige: *{high, medium, low}* |
| Fuel economy: *Number* | Fuel: *{petrol, diesel, battery}* | Number of models: *Ordinal* |
| Type: *{saloon, MPV, sports}* | Number of valves: *Ordinal* | Size: *{large, medium, small}* |
| Synonym: *automobile* | Synonym for 'Capacity': *size* | Synonym: *{make, producer}* |

Figure 6. Three examples of concept frames
that could form part of an ontology

Frames can also be used to represent relations and their properties. 3 examples of this are shown in Figure 7:

| has part | part of | manufactures |
|---|---|---|
| Inverse: *part of* | Inverse: *has part* | Inverse: *manufactured by* |
| LHS: *{car, car component}* | LHS: *car component* | LHS: *manufacturer* |
| RHS: *car component* | RHS: *{car, car component}* | RHS: *{car, car component}* |
| Type: *Transitive* | Type: *Transitive* | Type: *Anti-symmetric* |
| Synonym: *includes* | Synonym: *composed of* | Synonym: *makes* |

Figure 7. Three examples of relation frames
that could form part of an ontology

As some of this may be unclear to you, here is some explanation:

- Inverse means the relation that goes in the opposite direction, e.g. if $x$ – *has part* – $y$, then $y$ – *part of* – $x$;

- LHS means Left Hand Side, i.e. what class or classes of object can appear on the left of the relation when making a relationship triple;

- RHS means Right Hand Side, i.e. what class or classes of object can appear on the right of the relation when making a relationship triple;

- Transitive means that the same relation can 'jump' across relationships. For example, 'part of' is transitive, which means if $x – part of – y$, and $y – part of – z$, then $x – part of – z$.

Another way to show some of this information, is as a list of relationships (or triples) that can be formed by associating classes with relations, such as:

- car – has part – engine

- car – has part – wheel

- engine – has part – cam shaft

- manufacturer – manufactures – car

When the ontology is used, some of these general relationships will become instantiated with facts about cars, such as *Pontiac Bonneville – has part – V8 engine*, *Fiat – manufactures – Punto*.

As well as classes, attributes, relations and triples, the car ontology might also include rules and axioms, such as:

- Each car must have 1 and only 1 engine

- Each car must have 3 or more wheels

- A diesel engine must use diesel fuel

- If a car has two seats and has high acceleration then it is a sports car

- If a car has no engine then the car cannot be used as a vehicle

- If fuel economy is in km/hr then multiply by 5/8 to convert to mph

As you can see, the ontology specifies the knowledge in great detail and in ways that can be used by an IT system to make comparisons, manipulations and inferences.

## 5.3 How is an Ontology developed?

**KEYWORDS:**
SYSTEM DEVELOPMENT, TACIT KNOWLEDGE, TAXONOMY, ATTRIBUTES, FACETS

An ontology can be developed in a number of ways depending on factors such as:

- How much of the domain knowledge is already documented?

- How much of the domain knowledge is tacit knowledge lying deep inside the heads of experts?

- How much of the domain knowledge is already in a structured format?

- How many experts should be involved to create a common, shared view?

- How available are the domain experts to help build the ontology?

- What sort of ontology is required, and how will it be used?

Depending on the answers to these questions, a project might be approached in a number of ways:

- A knowledge engineer or ontological engineer builds the full ontology from scratch, including interviewing domain experts using the kind of techniques described in Section 4.3.2 and 4.3.3;

- A knowledge engineer or ontological engineer interviews a number of domain experts then builds a basic ontology structure which is then populated directly by domain experts;

- An existing ontology, or ontology framework, is used as the starting point, and then edited by the knowledge/ontological engineer and/or the domain experts;

A number of key decisions and activities must be undertaken, as described below.

### 5.3.1 What language and tool to use?

There are a number of languages available for representing the ontology, such as KIF, Ontolingua, Frame Logic, XOL and OWL. The decision on which to use will usually be based on the purpose of the ontology and the tools available. The choice of tool goes hand-in-hand with the choice of language, since certain tools can deal with one or two languages and not others. (See Section 5.5 for more information).

### 5.3.2 What to ask the domain experts to do?

How to involve domain experts will depend on a number of the factors listed at the beginning of this section. Obviously, the more experts are involved, the more of a common view can be created, although the process of integrating each person's way of conceptualising and describing the domain (e.g.

specific terminology) will take more time and be more prob-
lematic. Eliciting tacit knowledge, if needed, will require the
use of special techniques (see Section 4.3.3). If you can sit an
expert in front of a screen and have him/her enter parts of the
ontology, then this can generate good content as long as the
interface is user-friendly, the expert knows what to do and
can described knowledge in a structured way.

### 5.3.3 How to ensure the ontology fits its purpose?

A spiral approach is often the best way of ensuring an ontol-
ogy fits its purpose. This means that you perform develop-
ment cycles that get progressively longer and build progres-
sively bigger things. So, on the first cycle, you create a small
ontology, assess it, analyse the results and make the required
modifications. On the next cycle, you build more of the on-
tology, assess it, analyse the results and make the required
modifications. These cycles of ever-increasing complexity
end when the deliverable is complete.

### 5.3.4 What steps to take

Noy and McGuinness [13] have described a 7-step procedure
for developing an ontology: Step 1. Determine the domain
and scope of the ontology; Step 2. Consider reusing existing
ontologies; Step 3. Enumerate important terms in the ontolo-
gy; Step 4. Define the classes and the class hierarchy (taxon-
omy); Step 5. Define the properties of classes – slots (i.e. at-
tributes); Step 6. Define the facets of the slots (i.e. the sort of
values to use for each attribute); Step 7. Create instances.
These authors also suggest 3 pieces of general advice:

1. There is no one correct way to model a domain – there
   are always viable alternatives. The best solution almost
   always depends on the application that you have in
   mind and the extensions that you anticipate.

2. Ontology development is necessarily an iterative process.

3. Concepts in the ontology should be close to objects (physical or logical) and relationships in your domain of interest. These are most likely to be nouns (objects) or verbs (relationships) in sentences that describe your domain.

## 5.4 What are the uses of Ontologies?

### KEYWORDS:

INFORMATION INTEGRATION, GENERIC KNOWLEDGE, COMMONKADS, MOKA, GTO, WORDNET, CYC, TOVE, (KA)[2]

There are two main uses of ontologies: (i) As part of an IT system that can inspect, integrate, filter, manipulate and present information from different sources; (ii) As a generic knowledge structure when capturing and modelling knowledge.

### 5.4.1 Information Fusion, Filtering and Presentation

Ontologies have become important due to the massive explosion of information resources that has occurred over the past 20 years. The need has arisen for systems that can inspect and integrate information that is held in different locations with different formats and different structures. The role of an ontology in such a system is illustrated in Figure 8.

Let me describe how the system depicted in Figure 8 might operate. Suppose the user wishes to find the quickest and cheapest way to find a new engine for a car. The interface activates the web resources to look for a new engine. To do this it draws on information in the ontology.

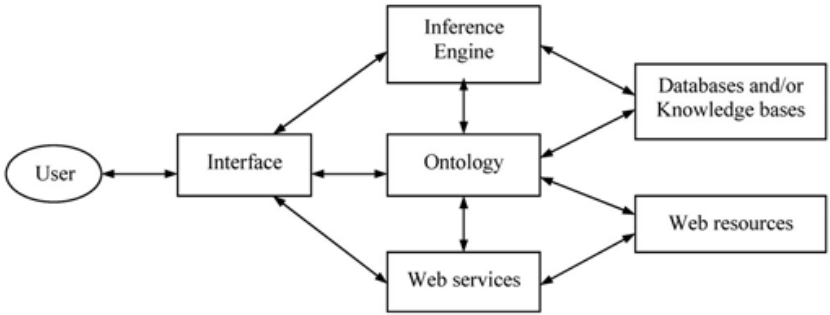

Figure 8. The role of an ontology in a semantic system

Different databases and web resources are checked for the availability of the engine and to gather information on costs, availability of stock, delivery dates, special fitment requirements etc. The inference engine is then tasked with creating a plan for a number of options then selecting the one that fits the user's needs. Again, the ontology is required to support the inference engine, giving information such as attributes and values of engines and constraints on the plan.

As this example shows, the ontology is acting as a kind of bridge that joins the other components together and allows them to understand each other (more details of which are given in the next chapter).

5.4.2 Generic Knowledge Structure

Ontologies can be used as the starting point for a knowledge model or knowledge base. Three examples of this are given below:

**CommonKADS** [3] uses a library of generic knowledge models (see Section 2.3.2). Each of these generic models describes a task type such as classification, diagnosis or plan-ning. It does so by providing two main elements: a problem solving model (showing the activities that need to be performed) and an ontology (showing the concepts involved, their relationships, attributes and axioms). For example, the ontology for diagnosis includes the concept 'hypothesis'. For a particular domain, this would be populated with specific knowledge, e.g. with diseases for a medical context, and with malfunctions for an engineering context.

**MOKA** [9] uses a framework for creating models of an engineering design process (see Section 3.3). For example MOKA include concepts such as entity and constraint to model the product being designed, and concepts such as activity and rule to model the design process. MOKA also defines the way these concepts link together and their attributes. This provides a template that can be filled-in when capturing knowledge of a specific design.

**GTO** (General Technological Ontology) [14] is a general template that extends the idea of MOKA to cover all aspects of an engineering or technology organisation. It includes 17 concepts, such as people, roles, knowledge areas, information resources, software, physical phenomena, functions, tasks and events. GTO has been used as the starting point for many knowledge bases that were used to create Knowledge Webs, particularly for Intranet systems (and not always in technological organisations).

5.4.3 Examples

To illustrate some uses of ontologies, here are 4 well-known ontologies:

- **WordNet**: This is a very large ontology of the English language containing around 100,000 words. Words are organised into sets having the same meaning (syno-nyms) with relations linking these sets, such as is-a and part-of. WordNet is available as a download and is free to use. It is mainly aimed at uses in linguistics.

- **Cyc**: This is a very large ontology of concepts aimed at modelling basic (common sense) knowledge required by intelligent computer systems. It contains hundreds of thousands of concepts with millions of logical axioms, rules and other assertions. The upper level of the Cyc taxonomy, containing 3000 concepts, is publicly available.

- **TOVE**: This is a comprehensive model of the knowledge in an organisation. It provides a shared terminology to be used by software agents and includes axioms to allow questions to be answered. It also defines a graphical means of depicting the concepts.

- **(KA)²**: This is an ontology of a specific area of research, namely knowledge acquisition. It comprises a number of ontologies, such as a project ontology, a person ontology, a research-topic ontology and a publications ontology. The ontology allows intelligent access to the knowledge.

## 5.5 What tools and technologies are there?

**KEYWORDS:**

LOGIC LANGUAGES, FRAME-BASED LANGUAGES, WEB LANGUAGES, PROTÉGÉ

In this section we take a brief look at ontology languages and ontology tools.

### 5.5.1 Languages

There are a number of ontology languages and formats available. Each one can be seen in terms of the following language types:

- **Logic languages**: Many ontologies are coded using logical expressions, i.e. formats developed by logicians. Logic is an area of Philosophy concerned with the representation and manipulation of statements so that their truth can be assessed. There are different types such as propositional logic, predicate logic and descriptive logic. For example, KIF and CycL are based on predicate logic; and the web language, OWL DL, is based on descriptive logic.

- **Frame-based Languages**: As the name suggests, these languages are based on frames (as was described in Section 5.2). For example, Ontolingua and Frame Logic use frame-based notations, as does the web language OIL (combined with descriptive logic).

- **Web Languages**: A number of languages have emerged in recent years for representing ontologies within web resources and web systems. All are forms of XML (described in Section 4.5.1). Early examples were XOL and OIL, then came DAML and DAML+OIL, and more recently RDF and OWL (see Section 6.5). Most of these are based on logic languages, with some also incorporating frame-based notations.

### 5.5.2 Tools

Fensel [11] describes a number of tools used for different aspects of ontologies:

- Tools for constructing and editing ontologies, such as Protégé (see below), OntoEdit, WebOnto, OilEd and ODE;

- Tools for reusing and merging ontologies such as SENSUS, Chimera, PROMPT, OntoMorph and OntoView;

- Tools for reasoning with ontologies, such as OntoBroker, SWI Prolog, CLIPS, Flora and FaCT;

- Tools for using ontologies to annotate the contents of an information resource;

- Tools for using ontologies to access and navigate an information resource, such as Ontobroker, On2broker and On-To-Knowledge.

Of all these tools the most widely used is **Protégé** [15]. Developed at Stanford University, Protégé has the advantage of being open source and freely available as a download. It is based on a technology (Java) that allows users to add software applications to the basic code and so enhance its functionality. Protégé uses a form-based editor, i.e. the user enters information for each class using a form that can be customised for a particular project. Users have developed add-ons that extend the functionality, e.g. graphical interfaces. Protégé supports both frame-based and logic-based languages.

Another tool that can be used to develop an ontology is PCPACK (see Section 2.5.1). This is a highly graphical tool that supports both a relation-based and a frame-based approach. It allows the user to create and edit OWL ontologies, which can be supplemented with less formal (human-readable) resources (as required in a Knowledge Web).

## *5.6 What are the issues of Ontologies?*

**KEYWORDS:**
ISSUES

### 5.6.1 Difficulty of Understanding

Many people new to ontologies can find them hard to understand. The reasons for this stem from the changing definitions (see later) and the language used to describe them. The much-quoted definition of an ontology as "a specification of a conceptualization" [12] adds to the confusion for some beginners. To address this issue, clear explanations need to be used, such as the one given on the W3C page [16]:

> "An ontology defines the terms used to describe and represent an area of knowledge. Ontologies are used by people, databases, and applications that need to share domain information (a domain is just a specific subject area or area of knowledge, like medicine, tool manufacturing, real estate, automobile repair, financial management, etc.). Ontologies include computer-usable definitions of basic concepts in the domain and the relationships among them".
> (from http://www.w3.org/TR/webont-req/)

Because many ontologies are used by software applications that do not understand the world (and have no "common sense"), the ontology has to be constructed in a very formal way. It also needs to be coded so that inferences can be made. To do this, most ontology languages are based on logic, which is a complex area that is difficult to understand by the uninitiated.

### 5.6.2 Changing Definition

Adding to the difficulty of understanding ontologies is the issue of a changing definition. Here is a brief history of the term.

- For most of its life, 'Ontology' has been an area of Philosophy concerned with the nature of existence. Central questions include: What kinds of objects exist? What is it for something to exist?

- In the 1980s, researchers in AI and Knowledge Engineering began to use the term to describe knowledge that was generic across a number of domains. In this sense, even a simple taxonomy was considered to be ontology, although other relationships were normally included.

- In the 1990s, the term began to be used by researchers in information science to mean "The hierarchical structuring of knowledge about things by subcategorising them according to their essential (or at least relevant and/or cognitive) qualities" [17].

- Since the 1990s, with the advent of the Semantic Web and semantic web technologies, the term is becoming more synonymous with anything that is coded in RDFS and OWL. A colleague working in the area has informed me that some people use the term for anything coded in OWL and for nothing else!

5.6.3 Generic or Specific

One of the technical issues debated in AI has been the generality of an ontology across different domains and applications. Ontologies are associated with the conceptual knowledge in a domain, whereas problem-solving models are associated with the procedural knowledge. It is debatable to what extent an ontology is independent of problem-solving knowledge (the interested reader will find some of this debate in the IJHCS special issue on ontologies [18]).

## *5.7 Where can I get more information?*

See references 11-18 in the Bibliography (Section 8).

Other sources of information:

# 6. SEMANTIC TECHNOLOGIES

## 6.1 What are Semantic Technologies?

**KEYWORDS:**
SEMANTIC TECHNOLOGIES, SEMANTIC WEB

Semantic Technologies are web technologies that provide sophisticated, knowledge-rich ways of storing and manipulating information. They allow web sites and other web resources to be understood and used by computers, as well as by humans.

Computer systems that can understand and manipulate web content are able to provide people with better services than a normal web site. A person is:

- Provided with personalised ways of accessing information;

- Able to select from different viewing options;

- Provided with solutions to tasks they would normally have to perform themselves;

- Able to view information that has been integrated from many different sources;

Semantic Technologies are very new. The trigger for their development was the Semantic Web, a vision for the next generation of the World Wide Web from its originator Sir Tim Berners-Lee. To quote from Sir Tim:

> "I have a dream for the Web [in which computers] become capable of analyzing all the data on the Web – the content, links, and transactions between people and computers. A 'Semantic Web', which should make this possible, has yet to emerge, but when it does, the day-to-day mechanisms of "trade, bureaucracy and our daily lives will be handled by machines talking to machines. The 'intelligent agents' people have touted for ages will finally materialize." [19].

To realise this vision in full would require major portions of the Web to be encoded in ways that allow computers to use them. There is a problem here – a contradiction that has hampered the development of a full-blown Semantic Web…

- To create the Semantic Web would mean constructing and converting millions (maybe billions) of web pages into machine-readable formats (e.g. RDF). For this to be achieved, without incurring massive investment, would require computers to read, understand and manipulate normal web content (i.e. written in normal sentences and paragraphs);

- Computers are not good at reading and understanding normal web content (i.e. sentences and paragraphs). That is the whole point of the Semantic Web – to encode the pages in formats that are machine-readable.

So the dream of replacing the World Wide Web with a Semantic Web is perhaps that, a dream. But this dream has spawned a new set of web technologies that can provide solutions to some of the major problems faced by organisations:

- How to get the best from the organisation's Intranet;

- How to integrate a large number of different databases and legacy systems without massive investment;

- How to improve the flow of information around the organisation;

- How to automate web-based tasks;

- How to provide decision-makers with relevant, up-to-date information in a format that is suited to their particular needs.

Semantic Technologies are beginning to provide solutions to these problems in many different domains, including medicine, engineering, research and military operations. The technologies involved are often quite light-weight and relatively inexpensive. As with all Knowledge Technologies, the emphasis is on providing adaptable and re-useable components. As Semantic Technologies develop and grow, they may become as ubiquitous and important as today's web browsers, search engines and web sites.

## *6.2 How does a Semantic System operate?*

**KEYWORDS:**

SEMANTIC SYSTEM, INFERENCE, TRIPLE STORES, ONTOLOGIES, WEB SERVICES

This section will describe a high-level picture of how a Semantic System operates. By 'Semantic System', I mean any system that combines:

- A triple store, i.e. a special type of knowledge base (see Section 6.5.3);

- An ontology or a number of ontologies (see Chapter 5);

- Web Services, i.e. web-based computer programs that can inspect, interpret and manipulate stored information (see Section 6.5.5);

- Traditional computer-based repositories of data and information, such as databases, electronic documents and web sites;

- A web-enabled interface for the human user.

Figure 9 shows the basic components and interactions of a typical Semantic System.

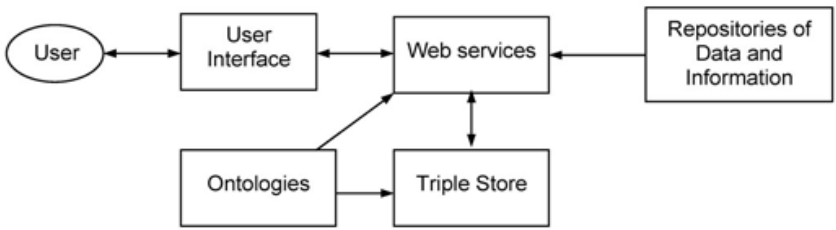

Figure 9. Components and interactions in a typical semantic system

How does such a system operate? There are two basic modes: a capture mode and a use mode. Let us look at some of the processes that occur in each of these modes.

6.2.1 Capture Mode

During capture mode the **web services** populate the triple store by inspecting and translating the relevant contents of **repositories**. To do this, the web services (combined with the triple store technologies) use the **ontologies** to understand the different schema and formats of each of the **repositories**. In this way the **triple store** becomes an integrated picture of the most relevant contents of the **repositories**.

6.2.2 Use Mode

During use mode, the operation might go something like this:

1. The user selects certain options on the user interface, e.g. wants to know something or wants the system to do something;

2. The relevant Web Services start to operate by inspecting and using the contents of the triple store;

3. Various inferences (see below) are made;

4. The triple store might show links to information in repositories that need to be retrieved;

5.  The web services and user interface pull all the rele-
    vant information together to provide the user with an
    answer, solution or picture of the situation.

## 6.2.3 Inference

Inference is an important part of a semantic system and all
Semantic Technologies are designed to allow inference to be
performed quickly on massive amounts of data and informa-
tion. Inference means creating a new piece of information
that is not present in the computer system. There are different
ways of doing this:

- **Inheritance**: This type of inference is used in frame-
  based systems. For example, imagine a system in
  which the class 'car' has an attribute 'number of
  wheels'. If I enter the value 4 for this attribute (denot-
  ing the fact that most cars have 4 wheels), then inherit-
  ance will mean that the value of 4 will become the de-
  fault value for all sub-classes and instances of 'car'. In
  other words, all types of cars will be given the proper-
  ty of 4 wheels. Any exceptions to this default value
  (e.g. 3-wheeler cars) can have the value altered after-
  wards.

- **Transitive relations**: A transitive relation is one such
  as 'faster than', 'part of' and 'is a' where the relation
  spans across relationships. For example, if 'A – faster
  than – B' and 'B – faster than – C' then it can be in-
  ferred that 'A – faster than – C'. So this is a form of
  inference that works across multiple relationships to
  create new relationships.

- **Class Definitions**: In some systems the membership of
  a class can be defined using rules, i.e. logical state-
  ments that define what can and cannot be a member of the class. These rules allow inference. For example, suppose I define the class membership of 'Faulty-Machine' with the rule '*member = Faulty-item AND Machine*'. So any member of the class 'Faulty-item' that is also a member of the class 'Machine' can be inferred to be a member of the class 'Faulty-Machine'.

- **Production Rules**: These are the types of rule used in many Knowledge Based Systems (as described in Section 2.2). They allow new information to be created by matching IF-THEN rules to a knowledge base or working memory.

Within a semantic system, Web Services and other computer programs would usually perform the inferential reasoning. Semantic languages such as RDF and OWL have been designed to allow inferences to be made and special technologies have been designed to do this. For example, Triple Stores are designed to include components that can perform inference on the triples to check for gaps and inconsistencies (see Section 6.5.3).

## 6.3 How is a Semantic System developed?

**KEYWORDS:**
SYSTEM DEVELOPMENT

Semantic systems differ in what they do and how they do it. The use of Knowledge Technologies in such systems is new and practices are evolving. As methodologies emerge for developing semantic systems, it is likely that many of the principles and practices used on more mature Knowledge Technologies (described in previous chapters) will be applicable.

Let us look at some of the activities that are relevant to the development of a semantic system:

1. You need to generate the content. This can be done in many ways. One way is the traditional way of authoring web pages by hand. Another way is to generate the content automatically using some sort of acquisition tool like PCPACK [4]. Another way is to work with existing databases, electronic documents or web sites.

2. You need to mark-up the content, i.e. add the tags that describe the semantic content. For that you need a mark-up language to describe the metadata, and the technology that is being used most successfully is RDF (see Section 6.5.2).

3. You need to create some sort of description of the domain behind the tagging so that the tagging refers to something. So it is useful to have an ontology (see Chapter 5) or even a knowledge base (see Section 1.4) that sits in the background. It should contain sufficient information to be useful to the system, e.g. when making smart links on the fly.

4. You need to select or write some software programs to manipulate and reason with the data and information. Web Services (see Section 6.5.5) are the main way of doing this, but any computer program could be used as long as it can interface to the web-based components in the rest of the system (see Section 6.5.6).

*6.4 What are the uses of Semantic Technologies?*

**KEYWORDS:**

SYSTEM INTEGRATION, TASK AUTOMATION, PROCESS IMPROVEMENT

Semantic Technologies are used when there is a need to:

- Provide decision-makers with relevant, up-to-date information;

- Improve the communication channels within an organisation, especially across functional boundaries;

- Automate time-consuming, web-based tasks;

- Integrate, filter and display information from a large number of different resources, such as databases, legacy systems and web sites;

- Provide the user with personalised ways of accessing and viewing information.

Let us examine some of these uses.

### 6.4.1 Decision-Making

Better decisions can be made when people have the right information at their finger tips. When information is out-of-date, incoherent, irrelevant, hard to find and hard to integrate, then the decision-making process is made more difficult. The information required to make good decisions often exists but tends to be scattered in various locations and stored in various formats. Knowledge Technologies provide ways of finding, fusing and exposing the information required to make informed decisions. One such system is MIAKT.

The MIAKT system [20] helps a medical team of different specialities (radiologists, histopathologists and clinicians) to appraise breast cancer cases. The software provides a number of functions for the team: (i) they can view and annotate various types of images, from x-ray mammograms to 3-dimensional MRI scans; (ii) they can search patient data; (iii) they can invoke services on the web for image analysis and data analysis.

### 6.4.2 Search and Retrieval

Traditional web search engines are useful but are not perfect. Everyone has had the experience of getting either no hits or 10,000 hits. Semantic Technologies make the process of search and retrieval smarter by understanding more about the context of the search. One such system is CAS (CS AKTive Space).

CAS [21] allows researchers in Computer Science to access all manner of web-based material. This is achieved using a Semantic Web of research material that provides multiple ways of looking at and discovering information and rich relations. For example, the user can select various filters to refine the queries that they make. In this way they can quickly home in on the information they require.

### 6.4.3 Integration and Presentation

We increasingly live and work within what people are calling an "Infosphere", i.e. an information rich environment. We need ways of using the mass of information and not being drowned in a sea of irrelevant and misleading information.

Organisations are awash with many different databases and other electronic resources. Finding inexpensive ways of integrating their contents, when required, is a major requirement. Ontologies allow Semantic Technologies to harvest the information and present it as an integrated picture. One such system is AKTiveSA.

AKTiveSA [22] uses a number of Semantic Technologies to provide military planners with improved situation awareness. The user interface is based on a map of the area with overlaid symbols and information. This provides an intuitive interface that requires little training. The interface includes features such as filtered visual displays, user alerts, naviga-tional and orientation aids, decision support services, certainty metrics, and so on.

### 6.4.4 Communication

Large organisations operate with thousands of people located in different places and having different specialist skills. Huge benefits can be gained by increasing the flow of information through the organisation. Semantic Technologies provide ways of improving this flow. To illustrate this, let us look at the IPAS system.

IPAS [23] is a system that enables design engineers to make more informed choices based on information from previous products, particularly when in service. IPAS focuses on the design, production and servicing of aero-engines. It allows designers to minimize maintenance costs using knowledge gained from maintenance histories of similar products. To do this, designers can inspect information that has been gathered from many different databases located in service centres all over the world.

## *6.5 What tools and technologies are available?*

**KEYWORDS:**
XML, RDF, RDFS, TRIPLE STORES, OWL, WEB SERVICES

This section describes the following Semantic Technologies: XML, RDF, RDFS, Triple Stores, OWL and Web services.

### 6.5.1 XML

As was described in Section 4.5.1, XML is a web mark-up language that allows users to define their own tags to encode information. XML is one of the main enabling technologies for semantic systems. As you will see later, the main semantic languages of RDF and OWL are both formed from XML. As was shown in section 4.5.2, an XML file can be searched for relevant information using an XSL stylesheet to present the information in different ways. This type of functionality forms the basis for many of the operations used in semantic systems.

### 6.5.2 RDF and RDFS

RDF (Resource Description Framework) is a standard format of XML for describing resources. A basic use of RDF is to describe metadata about a web resource, e.g. to describe information about a web page, such as who created it, when it was created, what it is about, what language it is in, and so on.

RDF can also describe the relationships between web resources, e.g. between 2 concepts in a knowledge base. As an illustration, the following piece of RDF contains 2 statements: the first says that a page in the PCPACK web site was created by Nick Milton; the second links the 'Nick Milton' concept to his email address:

```
<RDF …>
<Description
about="http:/www.pcpack.co.uk/Book/index.htm">
<DC:creator rdf:resource="www.xyzzz.uk/nickmilton/"/>
</Description>
<Description about="http:/www.xyzzz.uk/nickmilton/">
<DC:email
rdf:resource="nick.milton@tacitconnexions.com"/>
</Description>
</RDF>
```

In RDF, each statement is called a **triple** because it is formed of 3 parts: the **subject**, **predicate** and **object**. The subject will always be a knowledge resource (e.g. a web page) or a knowledge object (e.g. representing a person, an organisation or other entity). The predicate is a way of relating the subject and object, and is equivalent to a relation (e.g. is a, requires, owns, works for). So RDF is about coding knowledge in the form of triples such as:

- www.pcpack.co.uk/Book/ – creator – Nick Milton
- Nick Milton – email –nick.milton@tacitconnexions.com
- Nick Milton – livesIn – Nottingham
- Nottingham – partOf – UK
- UK – population – 60776238
- ukwmap.jpg – depicts – UK

A good way of visualising RDF is as a diagram called an RDF graph. An example of an RDF graph is shown in Figure 10.

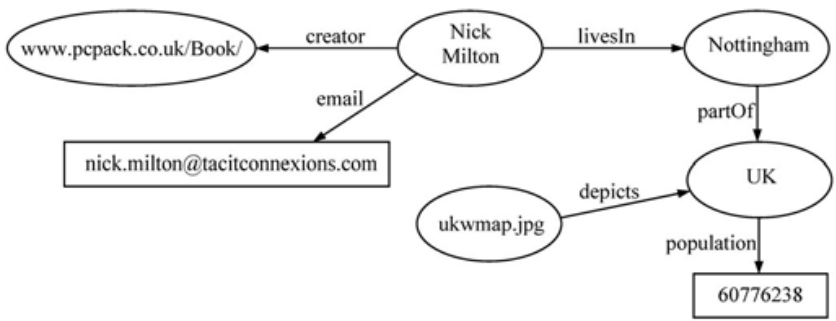

Figure 10. Example of an RDF graph

In RDF, the subject and predicate are identified by a URI (Uniform Resource Identifier). The URLs that we use to access a web site are one type of **URI**. Other URIs need not be addresses at all, even though they look like addresses. They are simply a way of identifying what something is. The ob-ject in a triple can be a URI or a **literal**, such as a number, a date or an email address.

To form more complex knowledge structures in RDF (e.g. "Steve Swallow knows that Nick Milton lives in Nottingham"), an RDF triple can be reified (made into a single knowledge object) so it can form the subject or object of a triple.

To define the RDF that is used for a particular domain, an **RDFS** (RDF Schema) is used. RDFS is a form of RDF that provides 3 core classes: Resource (to define the subjects), Property Type (to define the predicates) and Class (to define the objects). It also includes core property types such as *instanceOf*, *subClassOf* and *Constraint*.

### 6.5.3 Triple Stores

We saw in Section 6.2 that a triple store is a form of knowledge base that is used in a semantic system. It is a database that has been optimised for storing triples – millions of them. The contents of a triple store are not coded in RDF (or any other web language). Instead, the triple store is specialised form of relational database designed to handle millions of triples at very high speeds.

The format of a triple store has been deliberately made very simple so that huge amounts of data can be processed very rapidly using special software written to build and use triple stores. For example, triple stores in systems such as SESAME [24] or Oracle 11g [25] do not just hold information but can include user-defined rules that will perform inferences (see Section 6.2.3) on the triple store's contents.

The triple store in a semantic system is usually created and updated by compiling RDF to create the triples. Queries can be made on the triple store, or on the RDF, using a special language called SPARQL [26].

6.5.4 OWL

We saw above that RDF is good for building knowledge bases. RDF can be used to build ontologies but it is a clumsy approach, i.e. it is a lengthy and error-prone process. This is because ontologies require more sophisticated features than a knowledge base. Hence, the need for OWL (Web Ontology Language), a W3C standard for coding web-based ontologies.

OWL is a type of RDFS, hence it is a form of RDF, i.e. a form of XML. These languages are often depicted in a layered representation:

- XML is the base layer…

- then RDF is an XML dialect…

- then RDFS is a set of RDF constructs…

- then OWL is a form of RDFS for constructing ontologies.

Compared to RDFS, OWL provides a richer set of building blocks to construct knowledge objects, i.e. it has richer ways of describing object classes and object properties. It introduces some properties, and richer forms of relationships in terms of anonymous classes that can be defined by their properties or properties of other classes.

OWL follows a logic-based approach. It uses Description Logic (DL) for most of its constructs but there are constructs that are not represented in DL. This gives rise to three variants of OWL:

- **OWL DL**: Only includes the constructs of Descriptive Logic. This allows logical reasoning to prove or disprove things about the contents, spot contradictions and validate the ontology, e.g. with a tool like RACER [27];

- **OWL Full**: For people who want to use OWL as a set of vocabulary extensions to RDFS, without being able to prove anything;

- **OWL Lite**: For students learning to use OWL and so hides some of the complexities of DL.

At the moment, OWL DL is probably the most important, since most people are using OWL for the DL features. However, this can often limit the capacity for creating domain knowledge (to things that can be proved or disproved).

### 6.5.5 Web Services

Web services are one of the enabling technologies of Semantic Systems. They provide both the interface and the underlying logic. A web service may often 'reason' about the content of a web page and improve the content or it may build the content from scratch, but the knowledge to drive that usually comes from a separate knowledge base, as well as any 'template' web page.

An important technology for web services is SOAP (Simple Object Access Protocol) [28]. This is an XML-based standard for distributed computing, i.e. it helps web services to talk to one another. A SOAP package contains 2 parts: (i) the message, i.e. the data going to and from a web service, and (ii) an envelope, i.e. the technical information to say which web service is required, what parameters are expected, and so on.

### 6.5.6 Web Programming Environments

A Semantic System includes the code required to filter, manipulate, and present information. This could include some server side logic, which could be delivered through web services. But it need not; it could be delivered with web program-ming environments such as PHP, ASP.NET or JAVA. These environments all contain the necessary libraries to make it easy to use web services.

## 6.6 What are the issues of Semantic Technologies?

### 6.6.1 Theory and Practice

If you talk to some purists they might say that Semantic Webs are all about RDFS and Web Services. This view is fine for concentrating on what the systems can do, but software engineers and knowledge engineers who build real-world applications need to use other technologies to make the systems more efficient and scalable. Many other enabling technologies are being used, such as PHP and ASP.NET.

As for the current status of OWL as being the only web language to use of ontologies, many Semantic Web applications have been built using RDF and RDFS, and have not used OWL.

### 6.6.2 Scalability

One of the practical issues facing commercial uses of triple stores is scalability, i.e. how many triples can be feasibly stored and used? Two or three years ago, it was good if a triple store could store 10 million triples and could search quickly and get answers back in a sensible amount of time. But people are now talking about 100 billion triples. How do you make a triple store that is scalable to orders of magnitude increases in size? That is a different branch of Computer Science to producing the Semantic Webs, i.e. the actual building of triple stores that work well with massive amounts to information.

Triple stores are often constructed as distributed databases, and that is one of the ways in which the problem is solved, i.e. to break up the knowledge base into pieces. In this way, you might be able to apply the same solutions as applied to massive relational databases. Or maybe not, because triple stores are a lot more verbose than relational databases. This feature allows them to do clever things but it also means they can sometimes operate too slowly.

### 6.6.3 Open World

The same problems that apply to the Internet also apply to semantic systems. One of these is a so-called 'open world'. Just because you cannot find something does not mean it does not exist. A semantic system gives you more powerful mechanisms for finding things, and filtering things, but you can never associate the fact you cannot find something with the fact it does not exist, because a web is potentially infinitely large.

### 6.6.4 Identity Problems

Another issue to be dealt with is to do with identity problems, i.e. knowing exactly what you are talking about when the same name is used for many things, or where one thing can have many synonyms. This issue of identity has been given careful thought. It is the reason for Uniform Resource Identifiers (describe in Section 6.5.2) that provide identifiers that are both readable and allow uniqueness to be established quite easily.

6.6.5 Future of Semantic Technologies

More than any other of the Knowledge Technologies in this book, Semantic Technologies have the potential to provide a massive impact on the World Wide Web and on all our lives. There are many who think that Web 3 (the third incarnation of the Web) will be based mainly on Semantic Technologies. But not all agree with this. The future of technology and its mass use is hard, if not impossible, to foresee. It is highly likely that Semantic Technologies will increasingly become a major component of modern Intranet systems but whether they will take over the Internet is not so easy to see. What may be required is a killer application that will accelerate its impact and create the impetus for XML, RDF and OWL to rival the use of HTML as the major language of the Web.

## *6.7 Where can I get more information?*

See references 11 and 19-28 in the Bibliography (Section 8).

Other sources of information:

Berners-Lee,T., Hendler, J. and Lassila, O. (2001). The Semantic Web. *Scientific American*, May 2001, pp.34-43.

www.aktors.org/Publications/

http://www.w3.org/

# GLOSSARY

**AI (Artificial Intelligence)**. The discipline of building special computer systems that can perform complex activities usually only performed by humans.

**Attribute**. A quality or characteristic of a concept, e.g. weight, colour, usefulness - *see Section 1.4.1*.

**Class**. A group of concepts that are all of the same type.

**Concepts**. The individual items in a knowledge base that represent things such as physical entities, people, tasks, issues, documents, etc. - *see Section 1.4.1*.

**Conceptual Knowledge**. That part of expertise associated with the properties of concepts and the relationships between concepts. Also called 'declarative knowledge'.

**Description Logic**. A type of logical representation popular in semantic technologies (e.g. OWL) - *see Section 6.5.4*.

**Domain**. A subject area, discipline or topic, such as Thermodynamics or International Law.

**Domain Expert**. A person with substantial experience and expertise in a particular area, who is often the main source of knowledge when building a knowledge base or ontology. Also called 'expert', 'subject matter expert' or 'specialist'.

**Facet**. Term used by some ontological engineers for the features of a value to use for an attribute, e.g. the value type, the allowed values and the number of values.

**Frame**. A way of representing the attributes and values of a concept (class or instance) using a table - *see Section 1.4.2*.

**Inference**. A reasoning process that uses a set of rules or procedures to create new information - *see Section 6.2.3*.

**Instance**. A specific, unique concept that has no sub-types.

**Knowledge Base**. A special database that holds information representing the expertise of a particular domain - *see Section 1.4*.

**KBE**. Knowledge Based Engineering - *see Chapter 3*.

**Knowledge Based System**. A computer system that emulates the problem-solving capabilities of a human expert - *See Chapter 2*.

**Knowledge Book**. Term used in the KBE community for a Knowledge Web.

**Knowledge models**. Views of a knowledge base using diagrams and other structured representations, such as trees, maps, matrices and annotation pages.

**Knowledge Acquisition (KA)**. The activity of capturing, structuring and representing knowledge from any source for the purpose of storing, sharing or implementing the knowledge - *see Section 4.3*.

**Knowledge Elicitation**. The activity of capturing knowledge from a human expert.

**Knowledge Engineer**. The role of a person within a knowledge project who performs the knowledge acquisition and modelling and so creates a knowledge base or ontology.

**Knowledge Engineering**. The domain involved in creating knowledge bases for use in Knowledge Based Systems and other knowledge technologies.

**Knowledge Objects**. The elements that make up a knowledge base, e.g. concepts, relations, attributes and values - *see Section 1.4.1*.

**Knowledge Web**. A website created automatically from a knowledge base - *see Chapter 4*.

**Literal**. Term used in RDF for the object of a triple that represents a value such as a number, a categorical value (such as an adjective), a date, an email address, etc. - *see Section 6.5.2*.

**Ontology**. A specification of how knowledge will be represented - *see Chapter 5*.

**OWL**. Web Ontology Language - *see Section 6.5.4*.

**Predicate**. Term used in logic (and RDF) for a relation between 2 or more concepts.

**Procedural Knowledge**. The expertise required by a person or group of people to perform a complex process, task or activity.

**RDF**. Resource Description Framework - *see Section 6.5.2*.

**RDFS**. RDF Schema - *see Section 6.5.2*.

**Scoping**. The activity of selecting the specific areas of knowledge to be acquired during a knowledge project.

**Semantic Web**. A web that is used by computers - *see Chapter 6*.

**Tacit Knowledge**. Type of deep knowledge that is used to perform activities that seem to require no thought at all (at least no conscious thought).

**Taxonomy**. A hierarchical structure of concepts linked by 'is a' relations showing classes (categories) and their members.

**Triple**. A relationship between two concepts, e.g. 'book – written by – author'. So called, because there are three elements to the expression - *see Section 6.5.2*.

**Triple Store**. Special form of relational database for storing and handling triples - *see Section 6.5.3*.

**URI**. Uniform Resource Identifier - *see Section 6.5.2*.

**Value**. A specific quality or characteristic of a concept, e.g. heavy, red, useful - *see Section 1.4.1*.

**XML**. Extensible Mark-up Language - *see Section 4.5.1*.

**XSL**. XML Stylesheet Language - *see Section 4.5.2*.

# LIST OF KEYWORDS

(Some keywords appear duplicated in order to facilitate the search)

Creative Commons Legal Code
Attribution-NonCommercial 3.0 Unported

## License

THE WORK (AS DEFINED BELOW) IS PROVIDED UNDER THE TERMS OF THIS CREATIVE COMMONS PUBLIC LICENSE ("CCPL" OR "LICENSE"). THE WORK IS PROTECTED BY COPYRIGHT AND/OR OTHER APPLICABLE LAW. ANY USE OF THE WORK OTHER THAN AS AUTHORIZED UNDER THIS LICENSE OR COPYRIGHT LAW IS PROHIBITED.

BY EXERCISING ANY RIGHTS TO THE WORK PROVIDED HERE, YOU ACCEPT AND AGREE TO BE BOUND BY THE TERMS OF THIS LICENSE. TO THE EXTENT THIS LICENSE MAY BE CONSIDERED TO BE A CONTRACT, THE LICENSOR GRANTS YOU THE RIGHTS CONTAINED HERE IN CONSIDERATION OF YOUR ACCEPTANCE OF SUCH TERMS AND CONDITIONS.

## 1. Definitions

a) **"Adaptation"** means a work based upon the Work, or upon the Work and other pre-existing works, such as a translation, adaptation, derivative work, arrangement of music or other alterations of a literary or artistic work, or phonogram or performance and includes cinematographic adaptations or any other form in which the Work may be recast, transformed, or adapted including in any form recognizably derived from the original, except that a work that constitutes a Collection will not be considered an Adaptation for the purpose of this License. For the avoidance of doubt, where the Work is a musical work, performance or phonogram, the synchronization of the Work in timed-relation with a moving image ("synching") will be considered an Adaptation for the purpose of this License.

b) **"Collection"** means a collection of literary or artistic works, such as encyclopedias and anthologies, or performances, phonograms or broadcasts, or other works or subject matter other than works listed in Section 1(f) below, which, by reason of the selection and arrangement of their contents, constitute intellectual creations, in which the Work is included in its entirety in unmodified form along with one or more other contributions, each constituting separate and independent works in themselves, which together are assembled into a collective whole. A work that constitutes a Collection will not be considered an Adaptation (as defined above) for the purposes of this License.

c) "**Distribute**" means to make available to the public the original and copies of the Work or Adaptation, as appropriate, through sale or other transfer of ownership.

d) "**Licensor**" means the individual, individuals, entity or entities that offer(s) the Work under the terms of this License.

e) "**Original Author**" means, in the case of a literary or artistic work, the individual, individuals, entity or entities who created the Work or if no individual or entity can be identified, the publisher; and in addition (i) in the case of a performance the actors, singers, musicians, dancers, and other persons who act, sing, deliver, declaim, play in, interpret or otherwise perform literary or artistic works or expressions of folklore; (ii) in the case of a phonogram the producer being the person or legal entity who first fixes the sounds of a performance or other sounds; and, (iii) in the case of broadcasts, the organization that transmits the broadcast.

f) "**Work**" means the literary and/or artistic work offered under the terms of this License including without limitation any production in the literary, scientific and artistic domain, whatever may be the mode or form of its expression including digital form, such as a book, pamphlet and other writing; a lecture, address, sermon or other work of the same nature; a dramatic or dramatico-musical work; a choreographic work or entertainment in dumb show; a musical composition with or without words; a cinematographic work to which are assimilated works expressed by a process analogous to cinematography; a work of drawing, painting, architecture, sculpture, engraving or lithography; a photographic work to which are assimilated works expressed by a process analogous to photography; a work of applied art; an illustration, map, plan, sketch or three-dimensional work relative to geography, topography, architecture or science; a performance; a broadcast; a phonogram; a compilation of data to the extent it is protected as a copyrightable work; or a work performed by a variety or circus performer to the extent it is not otherwise considered a literary or artistic work.

g) "**You**" means an individual or entity exercising rights under this License who has not previously violated the terms of this License with respect to the Work, or who has received express permission from the Licensor to exercise rights under this License despite a previous violation.

h) "**Publicly Perform**" means to perform public recitations of the Work and to communicate to the public those public recitations, by any means or process, including by wire or wireless means or public digital performances; to make available to the public Works in such a way that members of the public may access these Works from a place and at a place individually chosen by them; to perform the Work to the public by any means or process and the communication to the public of the performances of the Work, including by public digital performance; to broadcast and rebroadcast the Work by any means including signs, sounds or images.

i) "**Reproduce**" means to make copies of the Work by any means including without limitation by sound or visual recordings and the right of fixation and reproducing fixations of the Work, including storage of a protected performance or phonogram in digital form or other electronic medium.

**2. Fair Dealing Rights.** Nothing in this License is intended to reduce, limit, or restrict any uses free from copyright or rights arising from limitations or exceptions that are provided for in connection with the copyright protection under copyright law or other applicable laws.

**3. License Grant.** Subject to the terms and conditions of this License, Licensor hereby grants You a worldwide, royalty-free, non-exclusive, perpetual (for the duration of the applicable copyright) license to exercise the rights in the Work as stated below:

a) to Reproduce the Work, to incorporate the Work into one or more Collections, and to Reproduce the Work as incorporated in the Collections;

b) to create and Reproduce Adaptations provided that any such Adaptation, including any translation in any medium, takes reasonable steps to clearly label, demarcate or otherwise identify that changes were made to the original Work. For example, a translation could be marked "The original work was translated from English to Spanish," or a modification could indicate "The original work has been modified.";

c) to Distribute and Publicly Perform the Work including as incorporated in Collections; and,

d) to Distribute and Publicly Perform Adaptations.

The above rights may be exercised in all media and formats whether now known or hereafter devised. The above rights include the right to make such modifications as are technically necessary to exercise the rights in other media and formats. Subject to Section 8(f), all rights not expressly granted by Licensor are hereby reserved, including but not limited to the rights set forth in Section 4(d).

**4. Restrictions.** The license granted in Section 3 above is expressly made subject to and limited by the following restrictions:

a) You may Distribute or Publicly Perform the Work only under the terms of this License. You must include a copy of, or the Uniform Resource Identifier (URI) for, this License with every copy of the Work You Distribute or Publicly Perform. You may not offer or impose any terms on the Work that restrict the terms of this License or the ability of the recipient of the Work to exercise the rights granted to that recipient under the terms of the License. You may not sublicense the Work. You must keep intact all notices that refer to this License and to the disclaimer of warranties with every copy of the Work You Distribute or Publicly Perform. When You Distribute or Publicly Perform the Work, You may not impose any effective technological measures on the Work that restrict the ability of a recipient of the Work from You to exercise the rights granted to that recipient under the terms of the License. This Section 4(a) applies to the Work as incorporated in a Collection, but this does not require the Collection apart from the Work itself to be made subject to the terms of this License. If You create a Collection, upon notice from any Licensor You must, to the extent practicable, remove from the Collection any credit as required by Section 4(c), as requested. If You create an Adaptation, upon notice from any Licensor You must, to the extent practicable, remove from the Adaptation any credit as required by Section 4(c), as requested.

b) You may not exercise any of the rights granted to You in Section 3 above in any manner that is primarily intended for or directed toward commercial advantage or private monetary compensation. The exchange of the Work for other copyrighted works by means of digital file-sharing or otherwise shall not be considered to be intended for or directed toward commercial advantage or private monetary compensation, provided there is no payment of any monetary compensation in connection with the exchange of copyrighted works.

c) If You Distribute, or Publicly Perform the Work or any Adaptations or Collections, You must, unless a request has been made pursuant to Section 4(a), keep intact all copyright notices for the Work and provide, reasonable to the medium or means You are utilizing: (i) the name of the Original Author (or pseudonym, if applicable) if supplied, and/or if the Original Author and/or Licensor designate another party or parties (e.g., a sponsor institute, publishing entity, journal) for attribution ("Attribution Parties") in Licensor's copyright notice, terms of service or by other reasonable means, the name of such party or parties; (ii) the title of the Work if supplied; (iii) to the extent reasonably practicable, the URI, if any, that Licensor specifies to be associated with the Work, unless such URI does not refer to the copyright notice or licensing information for the Work; and, (iv) consistent with Section 3(b), in the case of an Adaptation, a credit identifying the use of the Work in the Adaptation (e.g., "French translation of the Work by Original Author," or "Screenplay based on original Work by Original Author"). The credit required by this Section 4(c) may be implemented in any reasonable manner; provided, however, that in the case of a Adaptation or Collection, at a minimum such credit will appear, if a credit for all contributing authors of the Adaptation or Collection appears, then as part of these credits and in a manner at least as prominent as the credits for the other contributing authors. For the avoidance of doubt, You may only use the credit required by this Section for the purpose of attribution in the manner set out above and, by exercising Your rights under this License, You may not implicitly or explicitly assert or imply any connection with, sponsorship or endorsement by the Original Author, Licensor and/or Attribution Parties, as appropriate, of You or Your use of the Work, without the separate, express prior written permission of the Original Author, Licensor and/or Attribution Parties.

d) For the avoidance of doubt:

    i. i.Non-waivable Compulsory License Schemes. In those jurisdictions in which the right to collect royalties through any statutory or compulsory licensing scheme cannot be waived, the Licensor reserves the exclusive right to collect such royalties for any exercise by You of the rights granted under this License;

    ii. e.Waivable Compulsory License Schemes. In those jurisdictions in which the right to collect royalties through any statutory or compulsory licensing scheme can be waived, the Licensor reserves the exclusive right to collect such royalties for any exercise by You of the rights

granted under this License if Your exercise of such rights is for a purpose or use which is otherwise than noncommercial as permitted under Section 4(b) and otherwise waives the right to collect royalties through any statutory or compulsory licensing scheme; and,

iii. f.Voluntary License Schemes. The Licensor reserves the right to collect royalties, whether individually or, in the event that the Licensor is a member of a collecting society that administers voluntary licensing schemes, via that society, from any exercise by You of the rights granted under this License that is for a purpose or use which is otherwise than noncommercial as permitted under Section 4(c).

e) Except as otherwise agreed in writing by the Licensor or as may be otherwise permitted by applicable law, if You Reproduce, Distribute or Publicly Perform the Work either by itself or as part of any Adaptations or Collections, You must not distort, mutilate, modify or take other derogatory action in relation to the Work which would be prejudicial to the Original Author's honor or reputation. Licensor agrees that in those jurisdictions (e.g. Japan), in which any exercise of the right granted in Section 3(b) of this License (the right to make Adaptations) would be deemed to be a distortion, mutilation, modification or other derogatory action prejudicial to the Original Author's honor and reputation, the Licensor will waive or not assert, as appropriate, this Section, to the fullest extent permitted by the applicable national law, to enable You to reasonably exercise Your right under Section 3(b) of this License (right to make Adaptations) but not otherwise.

## 5. Representations, Warranties and Disclaimer

UNLESS OTHERWISE MUTUALLY AGREED TO BY THE PARTIES IN WRITING, LICENSOR OFFERS THE WORK AS-IS AND MAKES NO REPRESENTATIONS OR WARRANTIES OF ANY KIND CONCERNING THE WORK, EXPRESS, IMPLIED, STATUTORY OR OTHERWISE, INCLUDING, WITHOUT LIMITATION, WARRANTIES OF TITLE, MERCHANTIBILITY, FITNESS FOR A PARTICULAR PURPOSE, NONINFRINGEMENT, OR THE ABSENCE OF LATENT OR OTHER DEFECTS, ACCURACY, OR THE PRESENCE OF ABSENCE OF ERRORS, WHETHER OR NOT DISCOVERABLE. SOME JURISDICTIONS DO NOT ALLOW THE EXCLUSION OF IMPLIED WARRANTIES, SO SUCH EXCLUSION MAY NOT APPLY TO YOU.

**6. Limitation on Liability.** EXCEPT TO THE EXTENT REQUIRED BY APPLICABLE LAW, IN NO EVENT WILL LICENSOR BE LIABLE TO YOU ON ANY LEGAL THEORY FOR ANY SPECIAL, INCIDENTAL, CONSEQUENTIAL, PUNITIVE OR EXEMPLARY DAMAGES ARISING OUT OF THIS LICENSE OR THE USE OF THE WORK, EVEN IF LICENSOR HAS BEEN ADVISED OF THE POSSIBILITY OF SUCH DAMAGES.

## 7. Termination

a) This License and the rights granted hereunder will terminate automatically upon any breach by You of the terms of this License. Individuals or entities who have received Adaptations or Collections from You under this License, however, will not have their licenses terminated provided such individuals or entities remain in full compliance with those licenses. Sections 1, 2, 5, 6, 7, and 8 will survive any termination of this License.

b)   Subject to the above terms and conditions, the license granted here is perpetual (for the duration of the applicable copyright in the Work). Notwithstanding the above, Licensor reserves the right to release the Work under different license terms or to stop distributing the Work at any time; provided, however that any such election will not serve to withdraw this License (or any other license that has been, or is required to be, granted under the terms of this License), and this License will continue in full force and effect unless terminated as stated above.

## 8. Miscellaneous

a)   Each time You Distribute or Publicly Perform the Work or a Collection, the Licensor offers to the recipient a license to the Work on the same terms and conditions as the license granted to You under this License.

b)   Each time You Distribute or Publicly Perform an Adaptation, Licensor offers to the recipient a license to the original Work on the same terms and conditions as the license granted to You under this License.

c)   If any provision of this License is invalid or unenforceable under applicable law, it shall not affect the validity or enforceability of the remainder of the terms of this License, and without further action by the parties to this agreement, such provision shall be reformed to the minimum extent necessary to make such provision valid and enforceable.

d)   No term or provision of this License shall be deemed waived and no breach consented to unless such waiver or consent shall be in writing and signed by the party to be charged with such waiver or consent.

e)   This License constitutes the entire agreement between the parties with respect to the Work licensed here. There are no understandings, agreements or representations with respect to the Work not specified here. Licensor shall not be bound by any additional provisions that may appear in any communication from You. This License may not be modified without the mutual written agreement of the Licensor and You.

f)   The rights granted under, and the subject matter referenced, in this License were drafted utilizing the terminology of the Berne Convention for the Protection of Literary and Artistic Works (as amended on September 28, 1979), the Rome Convention of 1961, the WIPO Copyright Treaty of 1996, the WIPO Performances and Phonograms Treaty of 1996 and the Universal Copyright Convention (as revised on July 24, 1971). These rights and subject matter take effect in the relevant jurisdiction in which the License terms are sought to be enforced according to the corresponding provisions of the implementation of those treaty provisions in the applicable national law. If the standard suite of rights granted under applicable copyright law includes additional rights not granted under this License, such additional rights are deemed to be included in the License; this License is not intended to restrict the license of any rights under applicable law.

**Publishing studies** series

Several technologies are emerging that provide new ways to capture, store, present and use knowledge. This book is the first to provide a comprehensive introduction to five of the most important of these technologies: Knowledge Engineering, Knowledge Based Engineering, Knowledge Webs, Ontologies and Semantic Webs. For each of these, answers are given to a number of key questions (What is it? How does it operate? How is a system developed? What can it be used for? What tools are available? What are the main issues?). The book is aimed at students, researchers and practitioners interested in Knowledge Management, Artificial Intelligence, Design Engineering and Web Technologies.

During the 1990s, Nick worked at the University of Nottingham on the application of AI techniques to knowledge management and on various knowledge acquisition projects to develop expert systems for military applications. In 1999, he joined Epistemics where he worked on numerous knowledge projects and helped establish knowledge management programmes at large organisations in the engineering, technology and legal sectors. He is author of the book "Knowledge Acquisition in Practice", which describes a step-by-step procedure for acquiring and implementing expertise. He maintains strong links with leading research organisations working on knowledge technologies, such as knowledge-based engineering, ontologies and semantic technologies.

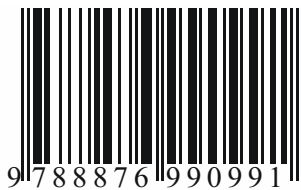

9 788876 990991